\documentclass[]{IEEEtran}
\usepackage{latexsym}
\usepackage{cite}
\usepackage{amssymb}
\usepackage{bbm}
\usepackage{bm}
\usepackage{mathtools}
\usepackage{stmaryrd}
\usepackage[usenames,dvipsnames]{color}
\usepackage{amsmath, amssymb,accents}
\usepackage{dsfont}
\usepackage{comment}
\usepackage[dvips]{graphics}
\DeclareMathOperator{\supp}{supp}
\usepackage{graphicx}
\usepackage{lipsum}
\usepackage{psfrag}
\usepackage{units}
\usepackage{setspace}
\usepackage{theorem}
\usepackage{float}
\usepackage{mathtools}

\allowdisplaybreaks

\newtheorem{definition}{Definition}

\newtheorem{theorem}{Theorem}
\newtheorem{lemma}{Lemma}
\newtheorem{remark}{Remark}

\newtheorem{corollary}{Corollary}
\newtheorem{proposition}{Proposition}

\newcommand\numberthis{\addtocounter{equation}{1}\tag{\theequation}}
{} 

\begin{document}

\makeatletter
\newcommand{\vasti}{\bBigg@{3}}
\newcommand{\vast}{\bBigg@{4}}
\newcommand{\Vast}{\bBigg@{5}}
\makeatother
\newcommand{\be}{\begin{equation}}
\newcommand{\ee}{\end{equation}}
\newcommand{\ba}{\begin{align}}
\newcommand{\ea}{\end{align}}
\newcommand{\baa}{\begin{align*}}
\newcommand{\eaa}{\end{align*}}
\newcommand{\bea}{\begin{eqnarray}}
\newcommand{\eea}{\end{eqnarray}}
\newcommand{\beaa}{\begin{eqnarray*}}
\newcommand{\eeaa}{\end{eqnarray*}}
\newcommand{\p}[1]{\left(#1\right)}
\newcommand{\pp}[1]{\left[#1\right]}
\newcommand{\ppp}[1]{\left\{#1\right\}}
\newcommand{\ber}{$\ \mbox{Ber}$}
\newcommand{\mkv}{{-\!\!\!\!\minuso\!\!\!\!-}}
\newcommand{\fillater}[1] {{\Large \color{red} #1}}

\title{Key and Message Semantic-Security over State-Dependent Channels}

\author{Alexander Bunin, Ziv Goldfeld, Haim H. Permuter, Shlomo Shamai (Shitz), Paul Cuff and Pablo Piantanida

\thanks{
Partial results of this work were presented at the 2017 international Workshop on Communication Security (WCS)~\cite{Bunin_WCS2017}, and at the 2018 IEEE International Symposium on Information Theory (ISIT).
}
\thanks{
An early version of this work was submitted to arXiv~\cite{Bunin_SkSm_Arxiv}. Proofs of several technical claims, which were omitted in this paper due to length restrictions, may be found in the appendices of the arXiv version.
}
\thanks{
The work of Alexander Bunin and Shlomo Shamai was supported by the European Union's Horizon 2020 Research and Innovation Programme, grant agreement No. 694630.
The work of Z. Goldfeld and H. H. Permuter was supported by the Israel Science Foundation (grant no. 684/11), an ERC starting grant and the Cyber Security Research Grant at Ben-Gurion University of the Negev.
Z. Goldfeld was also supported by the Rothschild Postdoc Fellowship and a grant from Skoltech--MIT Joint Next Generation Program (NGP).
The work of Paul Cuff was supported by the National Science Foundation, grant CCF-1350595, and the Air Force Office of Scientific Research, grant FA9550-15-1-0180.
}
\thanks{
A. Bunin and S. Shamai are with the Department of Electrical Engineering, Technion -- Israel Institute of Technology, Haifa, Israel (albun@tx.technion.ac.il,sshlomo@ee.technion.ac.il).
Z. Goldfeld is with the Department of Electrical Engineering and Computer Science, MIT, Cambridge, MA, USA (zivg@mit.edu). H. H. Permuter is with the Department of Electrical and Computer Engineering, Ben-Gurion University of the Negev, Beer-Sheva, Israel (haimp@bgu.ac.il).
P. Cuff is with the general research group at Renaissance Technologies, and was formerly with the Department of Electrical Engineering, Princeton University, Princeton, NJ, USA (cuff@princeton.edu).
P. Piantanida is with the Laboratory of Signals and Systems, CentraleSup\'elec-CNRS-Universit\'e, Paris-Sud, France (pablo.piantanida@centralesupelec.fr). }}

\maketitle


\begin{abstract}
    We study the trade-off between secret message (SM) and secret key (SK) rates, simultaneously achievable over a state-dependent (SD) wiretap channel (WTC) with non-causal channel state information (CSI) at the encoder. This model subsumes other instances of CSI availability as special cases, and calls for efficient utilization of the state sequence for both reliability and security purposes. An inner bound on the semantic-security (SS) SM-SK capacity region is derived based on a superposition coding scheme inspired by a past work of the authors. The region is shown to attain capacity for a certain class of SD-WTCs. SS is established by virtue of two versions of the strong soft-covering lemma. The derived region yields an improvement upon the previously best known SM-SK trade-off result reported by Prabhakaran \textit{et al.}, and, to the best of our knowledge, upon all other existing lower bounds for either SM or SK for this setup, even if the semantic security requirement is relaxed to weak secrecy. It is demonstrated that our region can be strictly larger than those reported in the preceding works.
\end{abstract}
	
\section{Introduction}\label{SEC:introduction}

    \subsection{Background}
	
	\IEEEPARstart{P}{hysical} layer security (PLS)~\cite{Bloch_Barros_Secrecy_Book2011,Liu_PhySecurity_Tutorial2017,Zeng2015_PLS_SK_Survey}, rooted in information-theoretic (IT) principles, is an approach to provably secure communication that dates back to Wyner's celebrated 1975 paper on the wiretap channel (WTC)~\cite{Wyner_Wiretap1975}. By harnessing randomness from the noisy communication channel and combining it with proper physical layer coding, PLS guarantees protection against computationally-unlimited eavesdroppers, with no requirement that the legitimate parties share a secret key (SK) in advance. Two fundamental questions in the field of PLS regard finding the best achievable transmission rate of a secret message (SM) over a noisy channel, and the highest attainable SK rate that distributed parties can agree upon based on correlated observations.
	
	The base model for SM transmission is Wyner's WTC~\cite{Wyner_Wiretap1975}, where two legitimate parties communicate over a noisy channel in the presence of an eavesdropper. The SM capacity of the degraded WTC was derived in~\cite{Wyner_Wiretap1975}, and the result was extended to the general case by Csisz{\'a}r and K{\"o}rner~\cite{Csiszar_Korner_BCconfidential1978}. The security analyses in~\cite{Wyner_Wiretap1975} and~\cite{Csiszar_Korner_BCconfidential1978} relied on evaluating particular conditional entropy terms, named \emph{equivocation}. This technique has been widely adopted in the IT community ever since. 
	
	Recently, distribution approximation arguments emerged as the tool of choice for proving security. This approach relies on a \emph{soft-covering lemma} (SCL) that originated in another 1975 paper by Wyner~\cite{Wyner_Common_Information1975}. The SCL states that the distribution induced by randomly selecting a codeword from an appropriately chosen codebook and passing it through a memoryless channel will be asymptotically indistinguishable from the distribution of random noise. The SCL was further developed over the years and stricter proximity measures between distributions were achieved~\cite{Han_Verdu_Resolvability1993,Kramer_resolvability2013,Goldfeld_WTCII2016,Goldfeld_AVWTC2016}. Based on these more advanced versions, one can make the channel output observed by the eavesdropper in the WTC seem like noise and, in particular, be approximately independent of the confidential data. This, in turn, implies IT security.
	Notably,~\cite{BastaniParizi2017SecrecyExponent} and~\cite{YagliCuff2018ISIT} focused on tight soft-covering exponents with respect to relative entropy and total variation, respectively.

	The study of SK agreement was pioneered by Maurer~\cite{Maurer1993}, and, independently, by Ahlswede and Csisz{\'a}r~\cite{AhlswedeCsiszar1993Part1}, who studied the achievable SK rates based on correlated observations at the terminals that can communicate via a noiseless and rate unlimited public link. The SK capacity when only one-way public communication is allowed was characterized in~\cite{AhlswedeCsiszar1993Part1}. This result was generalized in~\cite{csiszarNarayan2000CrHelper} to the case where the public link has finite capacity. The optimal random coding scheme for these cases is a combination of superposition coding and Wyner-Ziv coding~\cite{WynZiv76SideInformationDecoder}. If the encoder controls its source (rather than just observing it), this source becomes a channel input and the setup evolves to a WTC. This is a special case of the SK channel-type model that was also studied in~\cite{AhlswedeCsiszar1993Part1}.
	
	\subsection{Model and Contributions}

	A more general framework to consider is the state-dependent (SD) WTC with non-causal encoder channel state information (CSI). This model combines the WTC and the Gelfand and Pinsker (GP) channel~\cite{Gelfand_Pinsker}, and is therefore sometimes referred to as the GP-WTC. The dependence of the channel's transition probability on the state sequence accounts for the possible availability of correlated sources at the terminals. The similarity between the SM transmission and the SK agreement tasks makes their integration in a single model natural. Adhering to the most general framework, we study the SM-SK rate pairs that are simultaneously achievable over a SD-WTC with non-causal encoder CSI.
	
	
	\begin{figure}[!t]
		\begin{center}
			\begin{psfrags}
				\psfragscanon
				\psfrag{A}[][][.86]{\ \ \ \ \ $M$}
				\psfrag{B}[][][.86]{\ \ \ \ \ \ \ \ \ Encoder $f_n$}
				\psfrag{D}[][][.86]{\ \ \ \ \ \ \ \ \ \ \ \ }
				\psfrag{C}[][][.86]{\ \ \ \ $\mathbf{X}$}
				\psfrag{D}[][][.86]{$K$}
				\psfrag{E}[][][.86]{\ \ \ \ \ \ \ \ \ \ \ \ $W^n_{Y,Z|S,X}$}
				\psfrag{F}[][][.86]{\ \ \ \ $\mathbf{Y}$}
				\psfrag{G}[][][.86]{\ \ \ \ $\mathbf{Z}$}
				\psfrag{H}[][][.86]{\ \ \ \ \ \ \ \ \ Decoder $\phi_n$}
				\psfrag{I}[][][0.77]{\ \ \ \ \ \ \ \ \ \ Eavesdropper}
				\psfrag{J}[][][.86]{\ \ \ \ $(\hat{M},\hat{K})$}
				\psfrag{K}[][][0.69]{\ \ \ \ $\mspace{-1.72mu}M,\mspace{-0.86mu}K$}
				\psfrag{L}[][][.86]{\ \ \ \ $W^n_S$}
				\psfrag{M}[][][.86]{$\mathbf{S}$}
				\includegraphics[scale = .37]{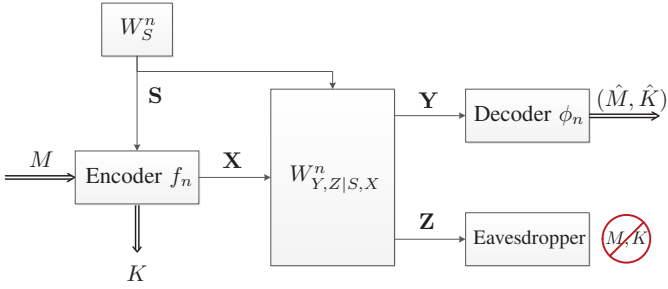}
				\caption{The state-dependent wiretap channel with non-casual encoder channel state information, exploited for simultaneous secret message transmission and secret key generation.} \label{FIG:wiretap}
				\psfragscanoff
				\vspace{-5mm}
			\end{psfrags}
		\end{center}
	\end{figure}

	The scenario where there is only a SM was studied in~\cite{SDWTC_Chen_HanVinck2006}, where an achievable SM rate formula was established. This result was improved in~\cite{Goldfeld_SDWTC2016} based on a novel superposition coding scheme\footnote{ The respective causal scenario was recently studied in~\cite{Fujita2016SDWTC_causalCSI,HanSaaki17SDWTC_causalCSI}.}.
	SK agreement over the GP-WTC was the focus of~\cite{Khisti_Key_Agreement2011}, and, more recently, of~\cite{Bassi2016secretKey} (see also references therein). The combined model was considered by Prabhakaran \emph{et al.}~\cite{Prabhakaran_SKSM2012}, who derived a benchmark inner bound on the SM-SK capacity region. The result from~\cite{Prabhakaran_SKSM2012} is optimal for several classes of SD-WTCs.
	
	We propose a superposition coding scheme for the combined model that subsumes all the aforementioned achievability results as special cases. Specifically,~\cite{Prabhakaran_SKSM2012,SDWTC_Chen_HanVinck2006,Goldfeld_SDWTC2016,Khisti_Key_Agreement2011,Bassi2016secretKey}, as well as all the other existing inner bounds (on SM transmission, SK agreement or both) that are known to the authors, are captured. Furthermore, our inner bound is shown to achieve strictly higher rates than each of these previous results. 
	
	The coding scheme used herein is an extension of the scheme in~\cite{Goldfeld_SDWTC2016}.
	Namely, an over-populated superposition codebook that encodes the entire confidential message in its outer layer is utilized.
	Using the redundancies in the inner and outer layers, the transmission is correlated with the state sequence by means of the likelihood encoder~\cite{Cuff_Song_Likelihood2016}.
	Constructing the inner codebook such that it is better observable by the eavesdropper (thus making the inner layer index decodable by him/her) enhances the secrecy resources that the legitimate parties can extract from the outer layer. The legitimate receiver decodes the entire codeword.

	Compared to the scheme from~\cite{Goldfeld_SDWTC2016}, and inspired by~\cite{Prabhakaran_SKSM2012}, our superposition code introduces an additional binning of the outer code layer (which also encodes the SM), that results in an additional redundancy index. Both redundancy indices are used to correlate the transmission with the observed state sequence. Based on distribution approximation arguments we show that the new index is approximately independent of the SM and uniform. Since the legitimate receiver decodes both layers, securing the new redundancy index along with the SM, establishes it as a SK.
	
	Our results are derived under the strict metric of semantic-security (SS). The SS criterion is a cryptographic gold standard that was adapted to the WTC framework (of computationally unbounded adversaries with a noisy observation) in~\cite{Vardy_Semantic_WTC2012}. As was shown in~\cite{Vardy_Semantic_WTC2012}, SS is equivalent to negligible mutual information (MI) between the confidential information (in our case, the SM-SK pair) and the eavesdropper's observations, when maximized over all possible message distributions. Our security analysis follows~\cite{Goldfeld_SDWTC2016}: the proof of SS relies on the strong SCL for superposition~\cite[Lemma 1]{Goldfeld_SDWTC2016} and the heterogeneous SCL~\cite[Lemma~1]{Goldfeld_AVWTC2016}.
	Since the past secrecy results from~\cite{SDWTC_Chen_HanVinck2006,Khisti_Key_Agreement2011,Bassi2016secretKey,Prabhakaran_SKSM2012} were derived under the weak secrecy metric (i.e., a vanishing \emph{normalized} MI with respect to a \emph{uniformly distributed} message-key pair), our achievability outperforms those schemes, not only in terms of the achievable rate pairs, but also in the upgraded sense of security.
	
	To conclude, the contribution of this work is as follows.
	We propose a coding scheme that generalizes~\cite{Goldfeld_SDWTC2016} and~\cite{Prabhakaran_SKSM2012}.
	The analysis follows~\cite{Goldfeld_SDWTC2016}, which, in turn, implies SS.
	Our result is shown to outperform~\cite{Goldfeld_SDWTC2016} for SK generation, and~\cite{Prabhakaran_SKSM2012} for SM transmission.
	The latter is done by introducing a specific example.
	Our achievable region is also shown to improve upon the previously best-known inner bound on the SK capacity~\cite{Bassi2016secretKey}.
	The proposed region is shown to be optimal for a certain class of SD-WTCs.
	Finally, we show that a recently reported inner bound on the SK capacity for this setup~\cite{Zibaeenejad2015WtSiPublicComm}, that seemingly achieves higher rates than the result herein, may, in certain cases, be unachievable.
	More specifically, a condition seems to be missing in the result of~\cite{Zibaeenejad2015WtSiPublicComm}. Adding the missing condition, it becomes a special case of the result herein.

	\subsection{Organization}
	
	This paper is organized as follows. Section~\ref{SEC:Preliminaries_Setup} establishes notation and definitions and sets up the SD-WTC problem. Section~\ref{SEC:main_result} states our main result -- an inner bound on the SM-SK optimal trade-off region. In Section~\ref{SEC:tight_capacity_results} our inner bound is shown to be tight for a certain class of channels. In Section~\ref{SEC:past_results} we discuss past results captured within the considered framework, and illustrate the improvement our result yields. The proof of the main result is the content of Section~\ref{SEC:proofs}. Finally, Section~\ref{SEC:summary} summarizes the main achievements and outlines the main insights emerging from this work.
	
\section{Preliminaries and Problem Set-Up} \label{SEC:Preliminaries_Setup}
	
	\subsection{Preliminaries}\label{SUBSEC:preliminaries}
	
	\par We use the following notations. As is customary, $\mathbb{N}$ is the set of natural numbers, while $\mathbb{R}$ are the reals. We further define $\mathbb{R}_+=\{x\in\mathbb{R}|x\geq 0\}$. Given two real numbers $a,b$, we denote by $[a\mspace{-3mu}:\mspace{-3mu}b]$ the set of integers $\big\{n\in\mathbb{N}\big| a\leq n \leq b \big\}$. Calligraphic letters denote sets, e.g., $\mathcal{X}$, while $|\mathcal{X}|$ stands for the cardinality of $\mathcal{X}$. $\mathcal{X}^n$ denotes the $n$-fold Cartesian product of $\mathcal{X}$. An element of $\mathcal{X}^n$ is denoted by $x^n=(x_1,x_2,\ldots,x_n)$; whenever the dimension $n$ is clear from the context, vectors (or sequences) are denoted by boldface letters, e.g., $\mathbf{x}$.

	Let $\big(\Omega,\mathfrak{F},\mathbb{P}\big)$ be a probability space, where $\Omega$ is the sample space, $\mathfrak{F}$ is the $\sigma$-algebra and $\mathbb{P}$ is the probability measure. Random variables over $\big(\Omega,\mathfrak{F},\mathbb{P}\big)$ are denoted by uppercase letters, e.g., $X$, with conventions for random vectors similar to those for deterministic sequences. The probability of an event $\mathcal{A}\in\mathfrak{F}$ is denoted by $\mathbb{P}(\mathcal{A})$, while $\mathbb{P}(\mathcal{A}\big|\mathcal{B}\mspace{2mu})$ denotes the conditional probability of $\mathcal{A}$ given $\mathcal{B}$. We use $\mathds{1}_\mathcal{A}$ to denote the indicator function of $\mathcal{A}\in\mathfrak{F}$. The set of all probability mass functions (PMFs) on a finite set $\mathcal{X}$ is denoted by $\mathcal{P}(\mathcal{X})$, i.e., 
	\begin{equation}
	    \mathcal{P}(\mathcal{X})=\bigg\{p:\mathcal{X}\to [0,1]\bigg|\sum_{x\in\mathcal{X}}p(x)=1\bigg\}.
	\end{equation}
	PMFs are denoted by letters such as $p$ or $q$, with a subscript that identifies the random variable and its possible conditioning. For example, for two discrete correlated random variables $X$ and $Y$ over the same probability space, we use $p_X$, $p_{X,Y}$ and $p_{X|Y}$ to denote, respectively, the marginal PMF of $X$, the joint PMF of $(X,Y)$ and the conditional PMF of $X$ given $Y$. In particular, $p_{X|Y}:\mathcal{Y}\to\mathcal{P}(\mathcal{X})$ represents the stochastic matrix whose elements are given by $p_{X|Y}(x|y)=\mathbb{P}\big(X=x|Y=y\big)$. Expressions such as $p_{X,Y}=p_Xp_{Y|X}$ are to be understood as $p_{X,Y}(x,y)=p_X(x)p_{Y|X}(y|x)$, for all $(x,y)\in\mathcal{X}\times\mathcal{Y}$. Accordingly, when three random variables $X$, $Y$ and $Z$ satisfy $p_{X|Y,Z}=p_{X|Y}$, they form a Markov chain, which is denoted by $X \mkv Y \mkv Z$.

    Any PMF $q\in\mathcal{P}(\mathcal{X})$ gives rise to a probability measure on $(\mathcal{X},2^{\mathcal{X}})$\footnote{Here $2^\mathcal{X}$ stands for the power set of $\mathcal{X}$.}, which we denote by $\mathbb{P}_q$; accordingly, $\mathbb{P}_q\big(\mathcal{A})=\sum_{x\in\mathcal{A}}q(x)$ for every $\mathcal{A}\subseteq\mathcal{X}$. We use $\mathbb{E}_q$ to denote an expectation taken with respect to $\mathbb{P}_q$. Similarly, we use $H_q$ and $I_q$ to indicate that an entropy or a mutual information term are calculated with respect to the PMF $q$. For a random vector $X^n$, if the entries of $X^n$ are drawn in an independent and identically distributed (i.i.d.) manner according to $p_X$, then for every $\mathbf{x}\in\mathcal{X}^n$ we have $p_{X^n}(\mathbf{x})=\prod_{i=1}^np_X(x_i)$ and we write $p_{X^n}(\mathbf{x})=p_X^n(\mathbf{x})$. Similarly, if for every $(\mathbf{x},\mathbf{y})\in\mathcal{X}^n\times\mathcal{Y}^n$ we have $p_{Y^n|X^n}(\mathbf{y}|\mathbf{x})=\prod_{i=1}^np_{Y|X}(y_i|x_i)$, then we write $p_{Y^n|X^n}(\mathbf{y}|\mathbf{x})=p_{Y|X}^n(\mathbf{y}|\mathbf{x})$. The conditional product PMF $p_{Y|X}^n$ given a specific sequence $\mathbf{x}\in\mathcal{X}^n$ is denoted by $p_{Y|X=\mathbf{x}}^n$.
	
	The empirical PMF $\nu_{\mathbf{x}}$ of a sequence $\mathbf{x}\in\mathcal{X}^n$ is $\nu_{\mathbf{x}}(x)\triangleq\frac{N(x|\mathbf{x})}{n}$, where $N(x|\mathbf{x})=\sum_{i=1}^n\mathds{1}_{\{x_i=x\}}$. We use $\mathcal{T}_\epsilon^{n}(p_X)$ to denote the set of letter-typical sequences of length $n$ with respect to the PMF $p_X$ and the non-negative number $\epsilon$, i.e., we have
	\begin{equation*}
		\mathcal{T}_\epsilon^{n}(p_X)=\Big\{\mathbf{x}\in\mathcal{X}^n\Big|\mspace{5mu}\big|\nu_{\mathbf{x}}(x)-p_X(x)\big|\leq\epsilon p_X(x),\ \forall x\in\mathcal{X}\Big\}.
	\end{equation*}

    \begin{definition}[Total Variation]
    	Let $(\Omega,\mathfrak{F})$ be a measurable space and $\mu$ and $\nu$ be two probability measures on that space. The total variation between $\mu$ and $\nu$ is
    	\begin{subequations}
    		\begin{equation}
    		||\mu-\nu||_{\mathsf{TV}}=\sup_{\mathcal{A}\in\mathfrak{F}}\big|\mu(\mathcal{A})-\nu(\mathcal{A})\big|.\label{EQ:total_variation_def}
    	\end{equation}
    	If the sample space $\Omega$ is countable, $p,q\in\mathcal{P}(\Omega)$ and $\mathbb{P}_p$ and $\mathbb{P}_q$ are the probability measures induced by $p$ and $q$, respectively, then \eqref{EQ:total_variation_def} reduces to
    	\begin{equation}
    		||\mathbb{P}_p-\mathbb{P}_q||_{\mathsf{TV}}=\frac{1}{2}\sum_{x\in\Omega}\big|p(x)-q(x)\big|\triangleq||p-q||_{\mathsf{TV}}.\label{EQ:total_variation_def_discrete}
    	\end{equation}
    	\end{subequations}
    \end{definition}

	\subsection{Problem Setup}\label{SUBSEC:setup}
	
	We study the SD-WTC with non-causal encoder CSI, for which we establish a novel achievable region of semantically secured message-key rate pairs.
	
	Let $\mathcal{S},\ \mathcal{X},\ \mathcal{Y}$ and $\mathcal{Z}$ be finite sets. The $\big(\mathcal{S},\mathcal{X},\mathcal{Y},\mathcal{Z},W_S,W_{Y,Z|S,X}\big)$ discrete and memoryless (DM) SD-WTC with non-causal encoder CSI is shown in Fig.~\ref{FIG:wiretap}. A state sequence $\mathbf{s}\in\mathcal{S}^n$ is sampled in an i.i.d. manner according to $W_S$ and revealed in a non-causal fashion to the sender. Independently of the observation of $\mathbf{s}$, the sender chooses a message $m$ from the set $\big[1:2^{nR_M}\big]$ and maps the pair $(\mathbf{s},m)$ onto a channel input sequence $\mathbf{x}\in\mathcal{X}^n$ and a key index $k\in\big[1:2^{nR_K}\big]$ (the mapping may be random). The sequence $\mathbf{x}$ is transmitted over the SD-WTC with transition probability $W_{Y,Z|S,X}:\mathcal{S}\times\mathcal{X}\to\mathcal{P}(\mathcal{Y}\times\mathcal{Z})$. The output sequences $\mathbf{y}\in\mathcal{Y}^n$ and $\mathbf{z}\in\mathcal{Z}^n$ are observed by the receiver and the eavesdropper, respectively. Based on $\mathbf{y}$, the receiver produces the pair $(\hat{m},\hat{k})$, its estimates of $(m,k)$. The eavesdropper tries to glean whatever it can about the message-key pair from $\mathbf{z}$.
	
	\begin{remark}[Most General Model]\label{REM:most_general_setting}
		The considered model is the most general instance of a SD-WTC with non-causal CSI known at some or all of the terminals.
		(See also~\cite[Section II.C]{Khisti_Key_Agreement2011} and references therein.)
		Seemingly, the broadest model one may consider is when the SD-WTC $W_{\tilde{Y},\tilde{Z}|S_t,X,S_r,S_e}$ is driven by a triple of correlated state random variables $(S_t,S_r,S_e)\sim W_{S_t,S_r,S_e}$, where $S_t$, $S_r$ and $S_e$ are known to the transmitter, the receiver and the eavesdropper, respectively. However, setting $S=S_t$, $Y=(\tilde{Y},S_r)$, $Z=(\tilde{Z},S_e)$ in a SD-WTC with non-causal encoder CSI and defining the channel's transition probability as
		\begin{equation*}
			W_{Y,Z|S,X}=W_{(\tilde{Y},S_r),(\tilde{Z},S_e)|S,X}=W_{S_r,S_e|S_t}W_{\tilde{Y},\tilde{Z}|S_t,X,S_r,S_e},
		\end{equation*}
		one recovers the aforementioned SD-WTC from the model with non-causal encoder CSI only. Our model also supports the existence of a \emph{public} or a \emph{private} bit-pipe (respectively, from the transmitter to the receiver and the eavesdropper, or to the receiver only), in addition to, or instead of, the noisy channel.
	\end{remark}

	\begin{definition}[Code]\label{DEF:SDWTC_code}
		An $(n,R_M,R_K)$-code $c_n$ for the SD-WTC with non-causal encoder CSI and a message set $\mathcal{M}_n\triangleq\big[1:2^{nR_M}\big]$ and a key set $\mathcal{K}_n\triangleq\big[1:2^{nR_K}\big]$ is a pair of functions $\left(f_n,\phi_n\right)$ such that
		\begin{enumerate}
	        \item $f_n:\mathcal{M}_n\times\mathcal{S}^n\to\mathcal{P}(\mathcal{K}_n\times\mathcal{X}^n)$ is a stochastic encoder.
	        \item $\phi_n:\mathcal{Y}^n\to\mathcal{M}_n\times\mathcal{K}_n$ is the decoding function.	   
	    \end{enumerate}
	\end{definition}

	For any message distribution $p_M\in\mathcal{P}(\mathcal{M}_n)$ and any $(n,R_M,R_K)$-code $c_n$, the induced joint PMF is
	\begin{align*}
	p^{(c_n)}&(\mathbf{s},m,k,\mathbf{x},\mathbf{y},\mathbf{z},\hat{m},\hat{k})\mspace{-3mu}= \mspace{-2mu}W_S^n(\mathbf{s})P_M(m) \numberthis\label{EQ:SDWTC_induced_PMF}\\
	&\times f_n(k,\mathbf{x}|m,\mathbf{s})W^n_{Y,Z|S,X}(\mathbf{y},\mathbf{z}|\mathbf{s},\mathbf{x})\mathds{1}\mspace{-3mu}_{\big\{(\hat{m},\hat{k})=\phi_n(\mathbf{y})\big\}}\mspace{-2mu}&.
	\end{align*}
	The probability measure induced by $p^{(c_n)}$ is $\mathbb{P}_{p^{(c_n)}}$. The performance of $c_n$ is evaluated in terms of its rate pair $(R_M,R_K)$, its maximal decoding error probability, the key uniformity and independence metric, and the SS-metric.
	
	\begin{definition}[Error Probability]\label{DEF:SDWTC_error_probability}
    	The error probability of an $(n,R_M,R_K)$-code $c_n$ is
    	\begin{subequations}
    	\begin{equation}
    	    e(c_n)\triangleq\max_{m\in\mathcal{M}_n}e_m(c_n),\label{EQ:error_probability}
    	\end{equation}
    	where for any $m\in\mathcal{M}_n$
    	\begin{align*}
    	    &e_m(c_n)\triangleq\mathbb{P}_{p^{(c_n)}}\Big(\big(\hat{M},\hat{K}\big)\neq(m,K)\Big| M=m\Big)\\
    	    &\quad=\mspace{-15mu}\sum\limits_{\substack{(\mathbf{s},\mathbf{x})\\\in\mathcal{S}^n\times\mathcal{X}^n}}\mspace{-15mu}W^n_S(\mathbf{s})f_n(k,\mathbf{x}|m,\mathbf{s})\mspace{-15mu}\sum\limits_{\substack{\mathbf{y}\in\mathcal{Y}^n:\\\phi_n(\mathbf{y})\neq (m,k)}}\mspace{-25mu}W_{Y|S,X}^n(\mathbf{y}|\mathbf{s},\mathbf{x}),\numberthis\label{EQ:error_probability_message}
    	\end{align*}
    	and subscript $p^{(c_n)}$ denotes that the underlying PMF is \eqref{EQ:SDWTC_induced_PMF}.	
    	\end{subequations}
	\end{definition}
	
	\begin{remark}[Operational Interpretation of the Error Prob.]
    	The error probability in \eqref{EQ:error_probability} is defined by maximizing \eqref{EQ:error_probability_message} over the set of messages $\mathcal{M}_n$. The maximization is only with respect to the message (rather than with respect to the SM-SK pair) because, while the choice of $M\sim p_M$ is independent of the code $c_n$, the distribution of the SK, $K$, and its estimate, $\hat{K}$, is induced by the code (see \eqref{EQ:SDWTC_induced_PMF}). A similar logic applies for the subsequent definition of the key uniformity and independence metric.
	\end{remark}
	
	\begin{definition}[Key Uniformity and Independence Metric]\label{DEF:SDWTC_key_uniformity}
    	The key uniformity and independence (of the message) metric under the $(n,R_M,R_K)$-code $c_n$ is
    	\begin{subequations}
    	\begin{equation}
    	\delta(c_n) \triangleq \max_{m\in\mathcal{M}_n} \delta_m(c_n),    
    	\end{equation}
    	where for any $m\in\mathcal{M}_n$
    	\begin{equation}
    	    \delta_m(c_n) \triangleq \big|\big|p^{(c_n)}_{K|M=m}- p^{(U)}_{\mathcal{K}_n}\big|\big|_\mathsf{TV}   
    	\end{equation}\label{EQ:delta_definition}
    	\end{subequations}
    	and $p^{(U)}_{{\mathcal{K}_n}}$ is the uniform PMF over $\mathcal{K}_n$.
	\end{definition}

    \begin{definition}[Information Leakage and SS Metric]
    	The information leakage to the eavesdropper under the $(n,R_M,R_K)$-code $c_n$ and the message PMF $p_{M}\in\mathcal{P}(\mathcal{M}_n)$ is 
    	$\ell(p_{M},c_n)\triangleq I_{p^{(c_n)}}(M,K;\mathbf{Z})$, where $I_{p^{(c_n)}}$ denotes that the MI is taken with respect to \eqref{EQ:SDWTC_induced_PMF}. The SS metric with respect to $c_n$ is
    	\begin{equation}
    	    \ell_\mathsf{Sem}(c_n)\triangleq\max_{p_M\in\mathcal{P}(\mathcal{M}_n)}\ell(p_M,c_n).
    	\end{equation}
	\end{definition}
	
    \begin{definition}[Achievability]\label{DEF:SDWTC_achievability}
    	A pair $(R_M,R_K)\in\mathbb{R}_+^2$ is called an achievable SS message-key rate pair for the SD-WTC with non-causal encoder CSI, if for every $\epsilon>0$ and sufficiently large $n$ there exists an $(n,R_M,R_K)$-code $c_n$ with
    	\begin{equation}
    	    \max\big\{e(c_n),\delta(c_n),\ell_\mathsf{Sem}(c_n)\big\}\leq\epsilon.\label{EQ:achievability_def}
    	\end{equation}
	\end{definition}

	\begin{definition}[SS-Capacity]
		The SS SM-SK capacity region $\mathcal{C}_{\mathsf{Sem}}$ of the SD-WTC with non-causal encoder CSI is the convex closure of the set of all achievable SS message-key rate pairs. The SM (SK) capacity is the supremum of all achievable SM (SK) rates.
	\end{definition}
	

\section{Main Result}\label{SEC:main_result}
	
The main result of this work is a novel inner bound on the SS SM-SK capacity region of the SD-WTC with non-causal encoder CSI. Our achievable region is at least as good as the best known achievability results for the considered problem, and is strictly larger in some cases. To state our main result, let $\mathcal{U}$ and $\mathcal{V}$ be finite sets and for any $q_{U,V,X|S}:\mathcal{S}\to\mathcal{P}(\mathcal{U}\times\mathcal{V}\times\mathcal{X})$ define $\mathcal{R}_\mathsf{A}\left(q_{U,V,X|S}\right)$ to be the region of all rate pairs $(R_M,R_K)\in\mathbb{R}_+^2$ satisfying
\begin{subequations}\label{EQ:SDWTC_alt_lower_bound_prob}\begin{align}
        R_M&\leq I(U,V;Y)-I(U,V;S),\label{EQ:SDWTC_alt_lower_bound_prob1}\\
		R_M+R_K&\leq I(V;Y|U)-I(V;Z|U), \label{EQ:SDWTC_alt_lower_bound_prob2}\\
		R_M+R_K&\leq I(U,V;Y)-I(V;Z|U)-I(U;S), \label{EQ:SDWTC_alt_lower_bound_prob3}
\end{align}\end{subequations}
where the MI terms are calculated with respect to the joint PMF $W_Sq_{U,V,X|S}W_{Y,Z|S,X}$, under which $(U,V) \mkv (S,X) \mkv (Y,Z)$ forms a Markov chain. 

\begin{theorem}[SS SM-SK Capacity Inner Bound]\label{TM:SDWTC_lower_bound}
	The following inclusion holds:
	\begin{equation}
		\mathcal{C}_{\mathsf{Sem}}\supseteq \mathcal{R}_\mathsf{A}\triangleq\bigcup_{q_{U,V,X|S}}\mathcal{R}_\mathsf{A}\left(q_{U,V,X|S}\right),\label{EQ:SDWTC_capacity_lower_bound}
	\end{equation}
	and one may restrict the cardinalities of $U$ and $V$ to $|\mathcal{U}|\leq |\mathcal{X}||\mathcal{S}| +5$ and $|\mathcal{V}|\leq |\mathcal{X}|^2|\mathcal{S}|^2+5|\mathcal{X}||\mathcal{S}|+3$.
\end{theorem}

The proof of Theorem~\ref{TM:SDWTC_lower_bound} is given in Section~\ref{SEC:proofs}, and is based on a secured superposition coding scheme. An over-populated two-layered superposition codebook is constructed (independently of the state sequence), in which the entire secret message is encoded in the \emph{outer layer}. Thus, no data is carried by the inner layer. The likelihood encoder~\cite{Cuff_Song_Likelihood2016} uses the redundancies in the inner and outer codebooks to correlate the transmitted codewords with the observed state sequence. Upon doing so, part of the correlation index from the outer layer is declared by the encoder as the key. The inner layer is designed to utilize the part of the channel which is better observable by the eavesdropper. This saturates the eavesdropper with redundant information and leaves him/her with insufficient resources to extract any information on the SM-SK pair from the outer layer. The legitimate decoder, on the other hand, decodes both layers of the codebook and declares the appropriate indices as the decoded message-key pair.

\begin{remark}[Interpretation of Theorem~\ref{TM:SDWTC_lower_bound}]\label{REM:mainResult_interpretation}
	To get some intuitive understanding of the result of Theorem~\ref{TM:SDWTC_lower_bound}, we examine $\mathcal{R}_\mathsf{A}(q_{U,V,X|S})$ from two different perspectives: when the joint PMF $W_Sq_{U,V,X|S}W_{Y,Z|S,X}$ is such that $I(U;Y)\geq I(U;S)$, and when the opposite inequality holds.
	
	If $I(U;Y)\geq I(U;S)$, the third rate bound in $\mathcal{R}_\mathsf{A}(q_{U,V,X|S})$ becomes redundant and the dominating bounds are
	\begin{subequations}
		\begin{align} 
			R_M&\leq I(U,V;Y)-I(U,V;S),\label{EQ:interpretation_case1_rb1}\\
			R_M+R_K&\leq I(V;Y|U)-I(V;Z|U).\label{EQ:interpretation_case1_rb2}
		\end{align}
	\end{subequations}
	The right-hand side (RHS) of \eqref{EQ:interpretation_case1_rb1} is the total rate of reliable (secured and unsecured) communication that our superposition codebook supports
	(inequalities \eqref{EQ:main_proof_approx_rate_bound2} and \eqref{EQ:main_proof_reliability_bound2}).
	This clearly bounds the rate of the SM that may be transmitted. For \eqref{EQ:interpretation_case1_rb2}, the MI difference on the RHS is the total rate of secrecy resources that are produced by the outer layer of the codebook
	(inequalities \eqref{EQ:main_proof_reliability_bound1} and \eqref{EQ:main_proof_SS_RB_final}).
	Since the security of our SM-SK pair comes entirely from that outer layer, this MI difference is an upper bound on the sum of rates.
	Notice that the reliability  \eqref{EQ:interpretation_case1_rb1} and the security \eqref{EQ:interpretation_case1_rb2} bounds are reminiscent of the original GP~\cite{Gelfand_Pinsker} and Csisz\'ar and K\"orner~\cite{Csiszar_Korner_BCconfidential1978} results, respectively.
	
	For the opposite case, if $I(U;Y)<I(U;S)$, then the second inequality in $\mathcal{R}_\mathsf{A}$ is inactive and we are left with
	\begin{subequations}
		\begin{align}
			R_M & \leq I(U,V;Y)-I(U,V;S),\label{EQ:interpretation_case2_rb1}\\
			R_M+R_K & \leq I(V;Y|U)-I(V;Z|U)\nonumber\\
			&\qquad\qquad\qquad-\big[I(U;S)-I(U;Y)\big].\label{EQ:interpretation_case2_rb2}
		\end{align}
	\end{subequations}
	While the interpretation of \eqref{EQ:interpretation_case2_rb1} remains as before, to understand \eqref{EQ:interpretation_case2_rb2} consider the following. Since $I(U;S)$ is approximately the rate of the inner codebook
	(inequality \eqref{EQ:main_proof_approx_rate_bound1}),
	$I(U;Y)<I(U;S)$ means that looking solely at the inner layer, the decoder lacks the resolution to decode it. However, the success of our communication protocol relies on the decoder reliably decoding both layers. Therefore, in this case, some of the rate from the outer layer is allocated to convey the inner layer index. Recalling that our security analysis is based on revealing the inner layer to the eavesdropper, this rate allocation effectively results in a loss of $I(U;S)-I(U;Y)$ in the secrecy resources of the outer layer, giving rise to the rate bound from \eqref{EQ:interpretation_case2_rb2}.
\end{remark}

\begin{remark}[Optimization Domain]
	It was shown in~\cite{Goldfeld_SDWTC2016} that when \(R_K = 0\), we may restrict the optimization in Theorem~\ref{TM:SDWTC_lower_bound} to joint PMFs \(q_{U,V,X|S}\) satisfying \(I(U;Y) \ge I(U;S)\) without inflicting any reduction in the achievable SM-rate.
	However, the proof from~\cite{Goldfeld_SDWTC2016} does not extend to the case when \(R_K > 0\).
	Currently, it remains unknown whether or not maximizing only over PMFs with \(I(U;Y) \ge I(U;S)\) is sufficient to exhaust $\mathcal{R}_\mathsf{A}$ when \(R_K>0\).
\end{remark}

\begin{remark}[Alternative Representations of $\mathcal{R}_\mathsf{A}$]
    By defining \(\tilde{V}=(U,V)\), we see that it suffices to restrict the maximization in \eqref{EQ:SDWTC_capacity_lower_bound} to joint PMFs that satisfy the Markov chain \(U \mkv \tilde{V} \mkv (S,X) \mkv (Y,Z)\).
    
    Regardless of that, the two bounds on \(R_M+R_K\) from \eqref{EQ:SDWTC_alt_lower_bound_prob2}-\eqref{EQ:SDWTC_alt_lower_bound_prob3} can be equivalently written as the single bound
    \begin{align*}
        R_M+R_K &\le I(U,V;Y) - I(U,V;Z) \\
        &\quad - \max\big\{I(U;Y),I(U;S)\big\} + I(U;Z). \numberthis
    \end{align*}
    In this form, it is evident that maximizing only over joint PMFs satisfying \(I(U;Z) \ge \max\big\{I(U;Y),I(U;S)\big\}\) attains optimality. Indeed, if the opposite inequality holds, one could always choose \(\tilde{V}=(U,V)\) and \(\tilde{U}=\emptyset\) to achieve higher rates.
\end{remark}

\begin{remark}[Cardinality Bounds]
	The cardinality bounds on the auxiliary random variables $U$ and $V$ in Theorem~\ref{TM:SDWTC_lower_bound} are established by standard application of the Eggleston-Fenchel-Carath{\'e}odory theorem~\cite[Theorem 18]{Eggleston_Convexity1958} twice. The details are omitted.
\end{remark}

\begin{remark}[Adaptation to the Rate-Equivocation]
	A confidential transmission of a SM requires channel resources for both reliability and security. The lesser of the two resources, therefore, limits the feasible transmission rates. The main focus of this paper is utilization of the residual secrecy resources that the SD-WTC offers. However, if secrecy is the lesser resource, the superior capability of the channel to support reliable communication may be utilized by considering a \emph{Rate-Equivocation} framework.

	Equivocation represents the \emph{portion} of the message that can be secured from the eavesdropper. (See~\cite{Csiszar_Korner_BCconfidential1978,Shamai_ITSecurity2009} for formal definitions.) The rate-equivocation framework enables communicating at rates higher than the SM capacity, as long as full secrecy is forfeited.
	
	By adaptation of the arguments from the proof of Theorem~\ref{TM:SDWTC_lower_bound} (see Section~\ref{SEC:proofs}), it naturally extends to an inner bound on the rate-equivocation region of the considered SD-WTC.
	The achievable rate-equivocation region is attained from \eqref{EQ:SDWTC_alt_lower_bound_prob} by substituting \(R_M\) in the left-hand side (LHS) of \eqref{EQ:SDWTC_alt_lower_bound_prob1} with the total reliable rate \(R\), and substituting \(R_M+R_K\) in the LHS of \eqref{EQ:SDWTC_alt_lower_bound_prob2} and \eqref{EQ:SDWTC_alt_lower_bound_prob3} with the equivocation rate \(R_E\). For more details see~\cite{Bunin_SkSm_Arxiv}.	
\end{remark}
	
\section{Tight Capacity Results} \label{SEC:tight_capacity_results}
	
\begin{figure}[!t]
	\begin{center}
		\begin{psfrags}
			\psfragscanon
		    \psfrag{A}[][][.86]{\ \ \ \ \ $M$}
			\psfrag{B}[][][.86]{\ \ \ \ \ \ \ \ \ Encoder $f_n$}
			\psfrag{D}[][][.86]{\ \ \ \ \ \ \ \ \ \ \ \ }
			\psfrag{C}[][][.86]{\ \ \ \ $\mathbf{X}$}
			\psfrag{D}[][][.86]{$K$}
			\psfrag{E}[][][.86]{\ \ \ \ \ \ \ \ \ \ \ \ $W^n_{Y,Z|S,X}$}
			\psfrag{F}[][][.86]{\ \ \ \ $\mathbf{Y}$}
			\psfrag{G}[][][.86]{\ \ \ \ $\mathbf{Z}$}
			\psfrag{H}[][][.86]{\ \ \ \ \ \ \ \ \ Decoder $\phi_n$}
			\psfrag{I}[][][0.77]{\ \ \ \ \ \ \ \ \ \ Eavesdropper}
			\psfrag{J}[][][.86]{\ \ \ \ $(\hat{M},\hat{K})$}
			\psfrag{K}[][][0.69]{\ \ \ \ $\mspace{-1.72mu}M,\mspace{-.86mu}K$}
			\psfrag{L}[][][.86]{\ \ \ \ $W^n_S$}
			\psfrag{M}[][][.86]{$\mathbf{S}$}
			\psfrag{N}[][][.86]{\ \ \ \ $W^n_L$}
			\psfrag{O}[][][.86]{$\mathbf{L}$}
			\includegraphics[scale = .37]{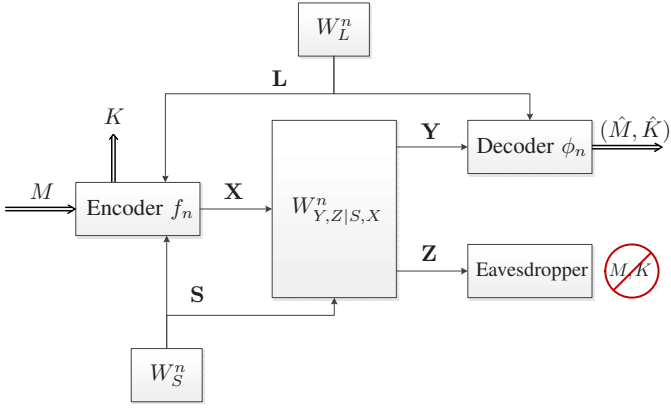}
			\caption{The SD less-noisy-eavesdropper WTC with a key.}\label{FIG:SWLNESK}
			\psfragscanoff
		\end{psfrags}\vspace{-6mm}
	\end{center}
\end{figure}

An operationally appealing special case of the considered SD-WTC is the following. Assume that $W_{Y,Z|S,X}$ is such that the eavesdropper's channel is less noisy than the main channel, but that the legitimate parties share a SK $\mathbf{L}\sim W_L^n$ (independent of the state sequence $\mathbf{S}\sim W_S^n$), using which they secure the confidential data. The setup is illustrated in Fig.~\ref{FIG:SWLNESK}. 

Formally, let $\mathcal{L}$, $\mathcal{S}$, $\mathcal{X}$, $\mathcal{Y}$ and $\mathcal{Z}$ be the alphabets of the key, the state, the channel input and the two channel outputs, respectively. The considered instance is the $\left(\tilde{\mathcal{S}},\mathcal{X},\tilde{\mathcal{Y}},\mathcal{Z}, W_{\tilde{S}}, W_{\tilde{Y},Z|\tilde{S},X}\right)$ SD-WTC with $\tilde{\mathcal{S}}=\mathcal{L}\times\mathcal{S}$, $\tilde{\mathcal{Y}}=\mathcal{L}\times\mathcal{Y}$,
$W_{\tilde{S}}=W_L\times W_S$, $\tilde{S}=(L,S)$, $\tilde{Y}=(L',Y)$, and whose channel transition matrix factors as
\begin{equation}
    W_{\tilde{Y},Z|\tilde{S},X}=W_{(L',Y),Z|(L,S),X}=\mathds{1}_{\{L'=L\}} W_{Y,Z|S,X},\label{EQ:SD_LNE_WTC_WK_channel}
\end{equation}
where $W_{Y,Z|S,X}$ is such that $Z$ is less noisy than $Y$. A less noisy $Z$ means that $I(U;Y)\leq I(U;Z)$ for any random variable $U$ for which $U\mkv(S,X)\mkv(Y,Z)$ forms a Markov chain. We refer to this special case as the \emph{SD less-noisy-eavesdropper WTC with a key}.


Theorem~\ref{TM:SDWTC_lower_bound} applies here since the above case is a certain instance of a SD-WTC with non-causal encoder CSI. As subsequently shown, the obtained inner bound is tight, thus characterizing the SS SM-SK capacity region of the SD less-noisy-eavesdropper WTC with a key. The following corollary states the result.

\begin{corollary}[SM-SK Capacity Region] \label{corol:EveHasBetterChannel_SK}
    The SS SM-SK capacity region of the SD less-noisy-eavesdropper WTC with a key is the set of all SM-SK rate pairs \((R_M,R_K) \in \mathbb{R}_+^2\) satisfying
    \begin{subequations}\label{EQ:EveHasBetterChannel_SK_bound}
    \begin{align}
        R_M &\le \max_{q_{U,X|S}} \left[I(U;Y) - I(U;S)\right], \label{EQ:EveHasBetterChannel_SK_boundReliability}\\
        R_K + R_M &\le H(L), \label{EQ:EveHasBetterChannel_SK_boundSecrecy}
    \end{align}
    \end{subequations}
    where the MI terms in \eqref{EQ:EveHasBetterChannel_SK_boundReliability} are with respect to the joint PMF $W_S q_{U,X|S} W_{Y|S,X}$.
\end{corollary}
The proof of Corollary~\ref{corol:EveHasBetterChannel_SK} is relegated to Appendix~\ref{APPEN:proof_corol_EveHasBetterChannel_SK}.
Note that while \eqref{EQ:EveHasBetterChannel_SK_boundReliability} bounds the total communication rate as a function only of the communication channel, \eqref{EQ:EveHasBetterChannel_SK_boundSecrecy} bounds the total secrecy rate depending solely on the secret source.

A direct consequence of Corollary~\ref{corol:EveHasBetterChannel_SK} is that when no SK is to be established between the legitimate parties, i.e., $R_K=0$, the best attainable SM rate is
\begin{equation}
    \mathcal{C}^{\mathsf{SM}} = \min\left\{ \max_{q_{U,X|S}} \big[ I(U;Y) - I(U;S) \big], H(L) \right\}.\label{EQ:SM_Capacity_Speical_Case}
\end{equation}
A simple separation-based coding scheme achieves the SM capacity from \eqref{EQ:SM_Capacity_Speical_Case}. Namely, using a capacity achieving error correction code, the channel is effectively converted into a reliable bit-pipe. Each of the legitimate parties compresses $\mathbf{L}$, which results in a uniform random variable. The latter is used to encrypt the SM via a one-time pad. The encrypted message is then transmitted over the reliable bit-pipe. Therefore, The achievable SM rate is equal to the minimum of the capacity of the channel $\max_{q_{U,X|S}} \big[ I(U;Y) - I(U;S) \big]$ and the rate of the key $H(L)$.

While this scheme may seem very natural, to the best of our knowledge, none of the past achievability results for the SD-WTC with non-causal CSI prior to~\cite{Goldfeld_SDWTC2016} attain its performance. In Section~\ref{subsubsec:prabh_exmpl}, a special case of this setup is used to demonstrate the improvement of our result over the previous benchmark achievable SM-SK region for the SD-WTC from~\cite{Prabhakaran_SKSM2012}.

\section{Previous Results as Special Cases}\label{SEC:past_results}
	
We compare the result of Theorem~\ref{TM:SDWTC_lower_bound} to those from related past works. The previously best known inner bound on the SM-SK trade-off region attainable over the considered SD-WTC is~\cite[Theorem 1]{Prabhakaran_SKSM2012}. The next subsection restates this inner bound and shows that Theorem~\ref{TM:SDWTC_lower_bound} can strictly outperform it. Afterwards, we provide a comparison to the best past achievability results for only SM transmission~\cite{Goldfeld_SDWTC2016} or only SK agreement~\cite{Bassi2016secretKey}.
The achievability result from~\cite{Goldfeld_SDWTC2016} captures the previous lower bounds on the SM capacity of the SD-WTC from~\cite{SDWTC_Chen_HanVinck2006,SDWTC_2Sided_Liu2007,SDWTC_Chia_ElGamal2012}. The SK achievability results from~\cite{Bassi2016secretKey} subsume previous lower bounds on the SK generation rate, such as~\cite{csiszarNarayan2000CrHelper,Khisti_Key_Agreement2011,Khisti_Key_Generation2012}.
Relating to one another these three benchmarks that we use to evaluate the performance of Theorem~\ref{TM:SDWTC_lower_bound}, we note that while~\cite{Goldfeld_SDWTC2016} recovers~\cite{Prabhakaran_SKSM2012} when there is only a SM ($R_K=0$),~\cite{Bassi2016secretKey} and~\cite{Prabhakaran_SKSM2012} do not imply one another.

It is noteworthy that many of the above mentioned achievability results were shown to be optimal for special instances of the studied model. Naturally, in all those cases, our result is optimal as well.

\begin{remark} \label{REM:zib_main}
    Another result on SK generation over SD-WTCs with non-causal CSI is found in~\cite{Zibaeenejad2015WtSiPublicComm}. Theorem 1 therein, which seemingly attains higher SK rates than both schemes from~\cite{Bassi2016secretKey} and our inner bound, is incorrect. The region suggested in~\cite[Theorem 1]{Zibaeenejad2015WtSiPublicComm}, in certain cases, exceeds the SK capacity, since it does not account for the loss in secrecy-rate when the inner layer codeword cannot be decoded on its own by the legitimate decoder, i.e., when $I(U;S)>I(U;Y)$. (See the second case in Remark~\ref{REM:mainResult_interpretation} for a further explanation.) For this reason, we chose~\cite{Bassi2016secretKey} as a benchmark for the SK generation problem.
    
    Looking at the proof of~\cite[Theorem 1]{Zibaeenejad2015WtSiPublicComm}, we conjecture that an additional constraint was assumed without being explicitly stated. Following the notations from~\cite{Zibaeenejad2015WtSiPublicComm}, the missing constraint seems to be
    \begin{equation}
        C_p + I(W;\check{Y}) > I(W;S),
    \end{equation}
    which would assure decodability of the inner code layer by the legitimate receiver without relying on the outer layer. Taking the additional constraint into consideration, our inner bound from Theorem~\ref{TM:SDWTC_lower_bound} recovers the amended Theorem 1 from~\cite{Zibaeenejad2015WtSiPublicComm} as follows.
    
    We use $(\tilde{U},\tilde{V},\tilde{X},\tilde{S},\tilde{Y},\tilde{Z})$ to denote the inner layer, the outer layer, the channel input, the encoder CSI, and the observations of the legitimate receiver and the eavesdropper, respectively, in Theorem 1 of~\cite{Zibaeenejad2015WtSiPublicComm}. These were originally denoted, respectively, by $W$, $U$, $X$, $S$, $\check{Y}$ and $\check{Z}$. To adjust our model to that of~\cite{Zibaeenejad2015WtSiPublicComm}, we identify $X=(\tilde{X},\Phi)$, $Y=(\tilde{Y},\Phi)$, $Z=(\tilde{Z},\Phi)$, $S=\tilde{S}$ in Theorem~\ref{TM:SDWTC_lower_bound}, where $\Phi$ is the random variable representing the input (and the outputs) of the public communication link. In order to comply with the rate restriction on the public link from~\cite{Zibaeenejad2015WtSiPublicComm}, we restrict the distribution of $\Phi$ to have $H(\Phi) \le C_P$.	Finally, we set:
    \begin{enumerate}
        \item $R_M=0$.
        \item $\Phi$ independent of $(\tilde{U},\tilde{V},\tilde{X},\tilde{S},\tilde{Y},\tilde{Z})$ with maximal entropy, i.e., such that $H(\Phi)=C_P$.
        \item $U=(\tilde{U},\Phi)$, $V=(\tilde{U},\tilde{V},\Phi)$.
    \end{enumerate}
    With respect to the above, substituting $(U,V,X,Y,Z,S)$ into \eqref{EQ:SDWTC_alt_lower_bound_prob} and maximizing only over distributions that satisfy $I(U;Y) - I(U;S) >0$ produces the amended version of~\cite[Theorem 1]{Zibaeenejad2015WtSiPublicComm}.
      
    To conclude the discussion of ~\cite[Theorem 1]{Zibaeenejad2015WtSiPublicComm} in its original form, a specific example showing the rates from that achievability formula to be exceeding the SK capacity is given in Appendix~\ref{APPEN:Zib}.
    We note that the missing condition in~\cite[Theorem 1]{Zibaeenejad2015WtSiPublicComm} does not seem to affect the correctness of the bulk of the other results therein.
\end{remark}

	
	\subsection{SM-SK Trade-off Region}\label{SUBSEC:Comparison_prabhakaran}
	
	The result of Theorem~\ref{TM:SDWTC_lower_bound} recovers the previously best known achievable SM-SK trade-off region over the SD-WTC with non-causal encoder CSI~\cite{Prabhakaran_SKSM2012}. In~\cite[Theorem 1]{Prabhakaran_SKSM2012} the following region was established:
	\begin{subequations}
		\begin{equation}
			\mathcal{R}_\mathsf{PER}\triangleq\bigcup_{q_U q_{V,X|U,S}}\mathcal{R}_\mathsf{PER}\left(q_U q_{V,X|U,S}\right),\label{EQ:compare_prabhakaran_LB}
		\end{equation}
		where, for any $q_U\in\mathcal{P}(\mathcal{U})$ and $q_{V,X|U,S}:\mathcal{U}\times\mathcal{S}\to\mathcal{P}(\mathcal{V}\times\mathcal{X})$,
		\begin{align*}
			&\mathcal{R}_\mathsf{PER}\left(q_U\times q_{V,X|U,S}\right)\triangleq\numberthis\label{EQ:compare_prabhakaran}\\
			&\left\{\mspace{-5mu}
			(R_M,R_K)\in\mathbb{R}_+^2\Bigg|\mspace{-7mu}\begin{array}{l}  R_M \leq I(U,V;Y)-I(U,V;S),\\
				R_M+R_K \leq I(V;Y|U)-I(V;Z|U)
			\end{array}\mspace{-10mu}\right\},
		\end{align*}\label{EQ:compaer_PBK_final}
		and the MI terms are taken with respect to $W_Sq_Uq_{V,X|U,S}W_{Y,Z|S,X}$, i.e., $U$ and $S$ are independent and $(U,V) \mkv (S,X) \mkv (Y,Z)$ forms a Markov chain.
	\end{subequations}
	
	First note that Theorem~\ref{TM:SDWTC_lower_bound} recovers $\mathcal{R}_\mathsf{PER}$ by restricting $U$ to be independent of $S$ in $\mathcal{R}_\mathsf{A}$. This is since for an independent pair $(U,S)$, we have $I(U;S)=0$, while $I(U,V;Y)\geq I(V;Y|U)$ always holds. Consequently, the third rate bound in $\mathcal{R}_\mathsf{A}$ becomes redundant and $\mathcal{R}_\mathsf{PER}$ is recovered.
	
	The result from~\cite{Prabhakaran_SKSM2012} was derived under the weak secrecy metric (i.e., a vanishing \emph{normalized} MI $\frac{1}{n}I(M,K;\mathbf{Z})$ between the SM-SK pair and the eavesdropper's observation sequence, where the message is assumed to be uniform). Our achievability, on the other hand, ensures SS. Theorem~\ref{TM:SDWTC_lower_bound}, therefore, improves upon~\cite[Theorem 1]{Prabhakaran_SKSM2012} both in the rates it achieves and in the sense of security it provides.

        \subsubsection{Achieving Strictly Higher Rates} \label{subsubsec:prabh_exmpl}
        
		Since~\cite[Theorem 1]{Prabhakaran_SKSM2012} allows only  inner layer random variables $U$ that are independent of the state, \emph{Gelfand-Pinsker} coding~\cite{Gelfand_Pinsker}, which generally requires correlating $U$ with $S$, is not supported in the inner layer.
		Instead, only \emph{Shannon's Strategies} coding~\cite{Shannon58}, which operates with independent $U$ and $S$ is allowed. The latter is optimal if the encoder observes the state \emph{causally}, but is generally sub-optimal when non-causal encoder CSI is available. To demonstrate the improvement of Theorem~\ref{TM:SDWTC_lower_bound} over~\cite{Prabhakaran_SKSM2012} we exploit the aforementioned limitation of the scheme therein, along with the observation that it is beneficial to exploit any part of a considered SD-WTC that is better observable by the eavesdropper to transmit the inner layer of the code.
		
		Let $\mathcal{X}=\mathcal{G}=\mathcal{L}=\mathcal{E}=\{0,1\}$, $\mathcal{S}=\{0,1,2\}$, $\mathcal{Y}=\{0,1,?\}$, where $? \notin \{0,1\}$ and $\mathcal{Z}=\mathcal{X}\times\mathcal{S}$. Consider the SD less-noisy-eavesdropper WTC with a key (defined in Section~\ref{SEC:tight_capacity_results}) shown in Fig.~\ref{FIG:prabh_exmpl_setup}, whose transition probability $W_{Y,Z|S,X}$, key $L\sim W_L$ and state $S\sim W_S$ are defined by the three parameters $\lambda,\epsilon,\sigma\in (0,0.5)$ as follows:
		
        \begin{figure}[t!]
        	\begin{center}
        		\begin{psfrags}
        			\psfragscanon
        			\psfrag{A}[][][.86]{\(M\)}
        			\psfrag{B}[][][.86]{\(\hat{M}\)}
        			\psfrag{C}[][][.86]{\(\mathbf{X}\)}
        			\psfrag{D}[][][.86]{\(\mathbf{S}\)}
        			\psfrag{E}[][][.86]{\(\mathbf{G}\)}
        			\psfrag{F}[][][.86]{\ \(\mathbf{Y}\)}
        			\psfrag{G}[][][.86]{\(\mathbf{L}\)~~}
        			\psfrag{H}[][][.86]{\(W^n_L\)}
        			\psfrag{I}[][][.86]{\(W^n_S\)}
        			\psfrag{J}[][][.86]{Encoder}
        			\psfrag{K}[][][.86]{$\begin{array}{c}
        					\text{Memory}\\ \text{with}\\\text{Stuck At}\\\text{Faults}
        				\end{array}$}
        			\psfrag{L}[][][.86]{$\mathsf{BEC}(\epsilon)$}
        			\psfrag{M}[][][.86]{Decoder}
        			\psfrag{N}[][][.86]{$\begin{array}{c}\text{Eaves-}\\\text{dropper}\end{array}$}
        			\includegraphics[scale = .35]{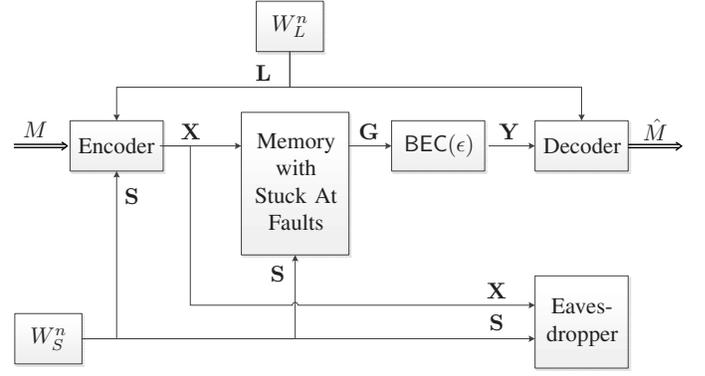}
        			\caption{Section~\ref{subsubsec:prabh_exmpl} example setup.} \label{FIG:prabh_exmpl_setup}
        			\vspace{-7mm}
        			\psfragscanoff
        		\end{psfrags}
        	\end{center}     
        \end{figure}

		\begin{itemize}	        
		    \item
		        $L$, $S$ and $E$ are independent random variables with $L\sim\mathsf{Ber}(\lambda)$, $E\sim\mathsf{Ber}(\epsilon)$ and 
                \begin{equation}
                    W_S(0)=W_S(1)=\frac{\sigma}{2}\quad ; \quad W_S(2)=1-\sigma.
                \end{equation}
            The joint distribution of $(L,S,E)$ is denoted by $W_{L,S,E}=W_L W_S W_E$.                
		    \item
		        The \emph{Memory with Stuck-at-Faults} (MSAF)~\cite{Tsybakov_Memory_stuckat1974} is a deterministic SD channel, driven by a ternary state $S$. The binary input and output symbols $X$ and $G$, respectively, are related through the function $g:\mathcal{S}\times\mathcal{X}\to\mathcal{G}$ given by
                \begin{equation}
                    g(s,x)=\begin{cases}
                    s,\quad s\in\{0,1\}\\
                    x,\quad s=2 
                    \end{cases}.\label{EQ:MSAF_def}
                \end{equation}
                
            \item
                The output of the MSAF channel is fed into a \emph{Binary Erasure Channel} with erasure probability $\epsilon$ (abbreviated as a $\mathsf{BEC}(\epsilon)$). The input $G$ and the ternary output $Y$ of the $\mathsf{BEC}(\epsilon)$ are related by means of the erasure random variable $E$ through the function $y:\mathcal{E}\times\mathcal{G}\to\mathcal{Y}$, where
                \begin{equation}
                    y(e,g)=\begin{cases}
                    g,\quad e=0\\
                    ?,\quad e=1 
                    \end{cases}.
                \end{equation}
                
            \item $Z=(S,X)$, i.e., the eavesdropper noiselessly observes the transmitted symbol $X$ and the state random variable $S$.

        \end{itemize}
    
        With respect to the above definitions, the transition matrix of the SD less-noisy-eavesdropper with channel $W_{Y,Z|S,X}$ is
        \begin{subequations}
        \begin{align*}
           &W_{Y,Z|S,X}(y',z|s,x)\\
           &\quad =\mspace{-10mu} \sum_{\substack{g' \in \{0,1\}\\e \in \{0,1\}}}W_E(e)W_{G,Y,Z|S,X,E}(g',y',z|s,x,e), \numberthis
        \end{align*}
        where
        \begin{equation}
            W_{G,Y,Z|S,X,E} =  \mathds{1}_{\{G=g(S,X)\}\cap\{Y=y(E,G)\}\cap\{Z=(S,X)\}}.
        \end{equation}
    	\end{subequations}
        A possible interpretation of this communication scenario is when the legitimate parties communicate through a public database that has memory faults known to the transmitter, but not to the receiver. The database and the faults are assumed to be known in full to the eavesdropper. To secure the communication the legitimate parties share a SK.
    			
    	For any $\lambda,\epsilon,\sigma\in(0,0.5)$, we denote the SM capacity of the corresponding channel by $C^{\mathsf{SM}}(\lambda,\epsilon,\sigma)$. Furthermore, let $R^{\mathsf{SM}}_\mathsf{A}(\lambda,\epsilon,\sigma)$ and $R^{\mathsf{SM}}_\mathsf{PER}(\lambda,\epsilon,\sigma)$ denote the maximal achievable SM rates attained by \eqref{EQ:SDWTC_capacity_lower_bound} from Theorem~\ref{TM:SDWTC_lower_bound} and \eqref{EQ:compare_prabhakaran} from~\cite[Theorem 1]{Prabhakaran_SKSM2012}, respectively. By virtue of Corollary~\ref{corol:EveHasBetterChannel_SK} (and, more specifically, \eqref{EQ:SM_Capacity_Speical_Case}), we have that Theorem~\ref{TM:SDWTC_lower_bound} is tight for the considered channel, i.e., 
    	\begin{equation}
    	    C^{\mathsf{SM}}(\lambda,\epsilon,\sigma)=R^{\mathsf{SM}}_\mathsf{A}(\lambda,\epsilon,\sigma),\quad \forall\lambda,\epsilon,\sigma\in(0,0.5).\label{EQ:we_achieve_capacity}
    	\end{equation}
    	As stated in the following proposition, $R^{\mathsf{SM}}_\mathsf{PER}(\lambda,\epsilon,\sigma)$ is strictly below capacity.	
    			
        \begin{proposition} \label{PROP:strict_improv_prbk}
    	    There exist $\lambda,\epsilon,\sigma\in(0,0.5)$ such that $R^{\mathsf{SM}}_\mathsf{PER}(\lambda,\epsilon,\sigma)<C^{\mathsf{SM}}(\lambda,\epsilon,\sigma)$.
    	\end{proposition}
    	
    	Proposition~\ref{PROP:strict_improv_prbk} is proven in Appendix~\ref{APPEN:proof_example_prbkrn}. The proof relies on the observation that for $R^{\mathsf{SM}}_\mathsf{PER}(\lambda,\epsilon,\sigma)$, a full utilization of the key $L$ implies that $R_M$ is upper bounded by the capacity of the considered channel with \emph{causal} CSI. In turn, this capacity is further upper bounded by the capacity of the MSAF with causal CSI. Choosing the parameters $\lambda,\epsilon,\sigma$ so that the SM capacity of the setup is strictly above the causal MSAF capacity, the superiority of our scheme compared to~\cite[Theorem 1]{Prabhakaran_SKSM2012} is established.

    	\begin{remark}
    	    This example actually demonstrates that~\cite[Theorem 1]{Goldfeld_SDWTC2016} (which is a special case of Theorem~\ref{TM:SDWTC_lower_bound}, when \(R_K=0\)) achieves strictly higher SM rates than~\cite[Theorem 1]{Prabhakaran_SKSM2012}.
    	\end{remark}

	
	\subsection{SM Transmission over SD-WTCs}\label{SUBSEC:Comparison_goldfeld}
	
	In~\cite[Theorem 1]{Goldfeld_SDWTC2016} a lower bound was established on the SS SM capacity (i.e., when $R_K=0$) over the considered SD-WTC. The SS SM capacity $C^{\mathsf{SM}}_\mathsf{Sem}$ was lower bounded by
	\begin{subequations}
		\begin{equation}
			C^{\mathsf{SM}}_\mathsf{Sem}\geq R_\mathsf{GCP}\triangleq\max_{q_{U,V,X|S}}R_\mathsf{GCP}\left(q_{U,V,X|S}\right),\label{EQ:compare_goldfeld_LB}
		\end{equation}
		where, for any $q_{U,V,X|S}:\mathcal{S}\to\mathcal{P}(\mathcal{U}\times\mathcal{V}\times\mathcal{X})$,
		\begin{align*}
			&R_\mathsf{GCP}\left(q_{U,V,X|S}\right)\\
			&\quad\triangleq\min\left\{\begin{array}{l}  I(U,V;Y)-I(U,V;S),\\
				I(V;Y|U)-I(V;Z|U),\\ I(U,V;Y)-I(V;Z|U)-I(U;S)
			\end{array}\right\},\numberthis\label{EQ:compare_goldfeld}
		\end{align*}
		and the MI terms are taken with respect to $W_Sq_{U,V,X|S}W_{Y,Z|S,X}$.
	\end{subequations}
	
	$R_\mathsf{GCP}$ is the projection in the $(R_M,R_K)$-plane of $\mathcal{R}_\mathsf{A}$ from Theorem~\ref{TM:SDWTC_lower_bound} to the $R_M$ axis when $R_K=0$. The main difference between the coding scheme from~\cite{Goldfeld_SDWTC2016} and our superposition code is the additional index $k\in\mathcal{K}_n$ in the outer layer of the codebook (which also encodes the SM $m\in\mathcal{M}_n$). Along with the other redundancy indices, $k$ is used to correlate the transmission with the observed state sequence via the likelihood encoder~\cite{Cuff_Song_Likelihood2016}. Based on distribution approximation arguments we show that $K$ is approximately independent of the message $M$ and approximately uniform. The pair $(M,K)$ is known to the transmitter and is reliably decoded by the receiver. Finally, by securing $K$ along with $M$ in our analysis, it is established as a SK.

	The intuition behind the SK construction is that, unlike the message, the key does not have to be independent of the state sequence, nor is it chosen by the user. Therefore, the redundancy index, used for correlating the codewords with the state sequence, is a valid key, as long as it is secured.

	Observing that any portion of the SM can be allocated in favor of a SK implies that \eqref{EQ:compare_goldfeld} is also an achievable SM-SK trade-off region, when $R_M$ above is replaced with $R_M+R_K$; however, this region is sub-optimal for SK generation. $\mathcal{R}_\mathsf{A}$ outperforms $R_\mathsf{GCP}$, e.g., in settings where an external random source $\mathbf{L}\sim W_L^n$ is observed by both legitimate parties but not by the eavesdropper, while the capacity of the communication channel is zero (say, $Y=Z=0$). For such a setup, the legitimate parties may use the random source to generate a SK of rate $H(L)$. While Theorem~\ref{TM:SDWTC_lower_bound} supports this strategy, $R_\mathsf{GCP}$ nullifies in this case. To see this, let $\tilde{S}\triangleq L$ and $\tilde{Y}\triangleq(L,Y)=(L,0)$ be the state and the channel output observed by the legitimate receiver, respectively. Inserting $\tilde{S}$ and $\tilde{Y}$ into the first term inside the minimum from \eqref{EQ:compare_goldfeld} produces $I(U,V;\tilde{Y})-I(U,V;\tilde{S})=0$, for any $q_{U,V,X|\tilde{S}}$.


	\subsection{SK Agreement over SD-WTCs}\label{SUBSEC:Comparison_bassi}
	
	In~\cite{Bassi2016secretKey} two achievable schemes were proposed for SK agreement over a WTC when the terminals have access to correlated sources. The results from~\cite{Bassi2016secretKey} do not imply one another. The difference between them is that~\cite[Theorem 2]{Bassi2016secretKey} is based on source and channel separation, while~\cite[Theorem 3]{Bassi2016secretKey} relies on joint coding.
	
	The setup in~\cite{Bassi2016secretKey} consists of three correlated sources $S_x$, $S_y$ and $S_z$ that are observed by the encoder, the decoder and the eavesdropper, respectively, and a SD-WTC in which the triple  $(S_x,S_y,S_z)$ plays the role of the state. Our general framework is defined through the state distribution $W_S$ and the SD-WTC $W_{\tilde{Y},\tilde{Z}|S,X}$. Setting $S=S_x$, $\tilde{Y}=(S_y,Y)$ and $\tilde{Z}=(S_z,Z)$ recovers the model from~\cite{Bassi2016secretKey} (see Remark~\ref{REM:most_general_setting}).
	
	The first scheme from~\cite[Theorem 2]{Bassi2016secretKey} operates under the assumption that the SD-WTC decomposes as $W_{(S_y,Y),(S_z,Z)|S_x,X}=W_{S_y,S_z|S_x}W_{Y,Z|X}$ into a product of two WTCs, one being independent of the state (given the input), while the other one depends only on it. Thus, the legitimate receiver (respectively, the eavesdropper) observes not only the output $\mathbf{Y}$ (respectively, $\mathbf{Z}$) of the WTC $W_{Y,Z|X}$, but also $\mathbf{S}_y$ (respectively, $\mathbf{S}_z$) - a noisy version of the state sequence drawn according to the corresponding conditional marginal of $W_{S_y,S_z|S_x}$. This scheme shows that the SK capacity $C^{\mathsf{SK}}$ is lower bounded by
	\begin{align*}
		C^{\mathsf{SK}}\geq R^{(\mathsf{Separate})}_\mathsf{BPS}\triangleq  &\max \Big[I(T;Y|Q)-I(T;Z|Q)\\
		&\quad+ I(\tilde{V};S_y|\tilde{U})-I(\tilde{V};S_z|\tilde{U})\Big],\numberthis\label{EQ:Bassi_separation}
	\end{align*}
	where the maximization is over all 
	$q_{\tilde{V}|S_x}q_{\tilde{U}|\tilde{V}}:\mathcal{S}_x\to\mathcal{P}(\tilde{\mathcal{V}}\times\tilde{\mathcal{U}})$ and $q_{Q,T}q_{X|T}\in\mathcal{P}(\mathcal{Q}\times\mathcal{T}\times\mathcal{X})$ that give rise to a joint PMF
	$W_{S_x,S_y,S_z} q_{\tilde{V}|S_x} q_{\tilde{U}|\tilde{V}} \times q_{Q,T}q_{X|T}W_{Y,Z|X}$ satisfying $I(\tilde{U};S_x|S_y)\leq I(Q;Y)$ and $I(\tilde{V};S_x|S_y)\leq I(T;Y)$.
	With respect to this distribution, $(S_y,S_z)\mkv S_x\mkv V\mkv U$ and $Q\mkv T\mkv X\mkv (Y,Z)$ form Markov chains and $(S_y,S_z,S_x,V,U)$ are independent of $(Q,T,X,Y,Z)$.
	This independence is the essence of separation that uses the channel for two purposes: carrying communication for SK agreement based on the sources, and securing part of this communication using wiretap coding.
	
	Setting $R_M=0$, $U=(Q,\tilde{U})$, $V=(T,\tilde{V})$ in Theorem~\ref{TM:SDWTC_lower_bound}, and limiting the union to joint PMFs that satisfy $I(U;S_y,Y) \ge I(U;S_x)$, recovers \eqref{EQ:Bassi_separation}. 
	
	The joint coding scheme from~\cite[Theorem 3]{Bassi2016secretKey} does not rely on the aforementioned decomposition of the SD-WTC $W_{(S_y,Y),(S_z,Z)|S,X_x}$. It lower bounds $C^{\mathsf{SK}}$ as
	\begin{equation}
		C^{\mathsf{SK}}\geq R^{(\mathsf{Joint})}_\mathsf{BPS}\triangleq\max \Big[I(\tilde{V};S_y,Y|\tilde{U})-I(\tilde{V};S_z,Z|\tilde{U})\Big]\label{EQ:Bassi_joint},
	\end{equation}
	where the maximization is over all 
	$q_{\tilde{V},X|S_x}q_{\tilde{U}|\tilde{V}}:\mathcal{S}_x\to\mathcal{P}(\tilde{\mathcal{V}}\times\mathcal{X}\times\tilde{\mathcal{U}})$ that give rise to a joint PMF \linebreak
	$W_{S_x} q_{\tilde{V},X|S_x} q_{\tilde{U}|\tilde{V}} W_{(S_y,Y),(S_z,Z)|S_x,X}$ satisfying $I(\tilde{U};S_x)\leq I(\tilde{U};S_y,Y)$ and $I(\tilde{V};S_x|\tilde{U})\leq I(\tilde{V};S_y,Y|\tilde{U})$. Setting $R_M=0$ and $(U,V)=(\tilde{U},\tilde{V})$ in Theorem~\ref{TM:SDWTC_lower_bound}, where $(\tilde{U},\tilde{V})$ is a valid auxiliary pair for
	$R^{(\mathsf{Joint})}_\mathsf{BPS}$, recovers \eqref{EQ:Bassi_joint}.
	
	It was shown in~\cite{Bassi2016secretKey} that, in some cases, the separation-based scheme achieves strictly higher rates than the joint coding scheme, i.e., that $R^{(\mathsf{Separate})}_\mathsf{BPS}>R^{(\mathsf{Joint})}_\mathsf{BPS}$. As Theorem~\ref{TM:SDWTC_lower_bound} captures both these results, it unifies the two schemes from~\cite{Bassi2016secretKey}, and, in particular, outperforms $R^{(\mathsf{Joint})}_\mathsf{BPS}$. Since the results from~\cite{Bassi2016secretKey} were derived under the weak secrecy metric, Theorem~\ref{TM:SDWTC_lower_bound} also upgrades them to SS (which is equivalent to strong secrecy when only SK generation is of interest).

\section{Proof of Theorem~\ref{TM:SDWTC_lower_bound}}\label{SEC:proofs}
	    
The subsequently presented proof follows lines similar to those from the proof of~\cite[Theorem 1]{Goldfeld_SDWTC2016}. Several claims herein are recovered from corresponding assertions in~\cite{Goldfeld_SDWTC2016} by identifying the index $j$ in~\cite{Goldfeld_SDWTC2016} with the pair $(j,k)$ in our scheme. The proofs of such claims are omitted, and the reader is referred to~\cite{Goldfeld_SDWTC2016}.

Fix $\epsilon>0$ and a conditional PMF $q_{U,V,X|S}:\mathcal{S}\to\mathcal{P}(\mathcal{U}\times\mathcal{V}\times\mathcal{X})$. For any $n\in\mathbb{N}$, let $p_M\in\mathcal{P}(\mathcal{M}_n)$ be the message distribution. We first show that for any $(R_M,R_K)\in\mathcal{R}_\mathsf{A}\left(q_{U,V,X|S}\right)$ there exists a SS sequence of $(n,R_M,R_K)$-codes with a key distribution that is approximately uniform conditioned on any message, and a vanishing \emph{average} error probability. We then use the expurgation technique~\cite[Theorem 7.7.1]{CovThom06} to ensure a vanishing \emph{maximal} error probability. This is done without harming the SS and the statistical properties of the key, since they hold for each message in the original message set.


\par\textbf{Codebook $\bm{\mathsf{B}_n}$:} We use a superposition codebook where the outer layer carries both the SM and the SK. The codebook is constructed independently of $\mathbf{S}$, but has sufficient redundancy to enable correlating the transmission with it.

Define the index sets $\mathcal{I}_n\triangleq\big[1:2^{nR_1}\big]$ and $\mathcal{J}_n\triangleq\big[1:2^{nR_2}\big]$. Let $\mathsf{B}_U^{(n)}\triangleq\big\{\mathbf{U}(i)\big\}_{i\in\mathcal{I}_n}$ be a random inner layer codebook, which is a set of random vectors of length $n$ that are i.i.d. according to $q_U^n$. An outcome of $\mathsf{B}_U^{(n)}$ is denoted by $\mathcal{B}_U^{(n)}\triangleq\big\{\mathbf{u}(i)\big\}_{i\in\mathcal{I}_n}$. 

To describe the outer layer codebook, fix $\mathcal{B}_U^{(n)}$ and, for every $i\in\mathcal{I}_n$ let $\mathsf{B}_V^{(n)}(i)\triangleq\big\{\mathbf{V}(i,j,k,m)\big\}_{(j,k,m)\in\mathcal{J}_n\times\mathcal{K}_n\times\mathcal{M}_n}$ be a collection of i.i.d. random vectors of length $n$ with distribution $q^n_{V|U=\mathbf{u}(i)}$. For each $i\in\mathcal{I}_n$, an outcome of $\mathsf{B}_V^{(n)}(i)$ given $\mathcal{B}_U^{(n)}$ is denoted by $\mathcal{B}_V^{(n)}(i)\triangleq\big\{\mathbf{v}(i,j,k,m)\big\}_{(j,k,m)\in\mathcal{J}_n\times\mathcal{K}_n\times\mathcal{M}_n}$. We also set $\mathsf{B}_V=\big\{\mathsf{B}_V(i)\big\}_{i\in\mathcal{I}_n}$ and denote its realizations by $\mathcal{B}_V$. Finally, a random superposition codebook is given by   $\mathsf{B}_n=\Big\{\mathsf{B}_U^{(n)},\mathsf{B}_V^{(n)}\Big\}$, while $\mathcal{B}_n=\Big\{\mathcal{B}_U^{(n)},\mathcal{B}_V^{(n)}\Big\}$ denotes a fixed codebook.

Let $\mathfrak{B}_n$ be the set of all possible outcomes of $\mathsf{B}_n$. The above codebook construction induces a PMF $\mu\in\mathcal{P}(\mathfrak{B}_n)$ over the codebook ensemble. For every $\mathcal{B}_n\in\mathfrak{B}_n$, we have
\begin{equation}
    \mu(\mathcal{B}_n)= \prod_{i\in\mathcal{I}_n}q^n_U\big(\mathbf{u}(i)\big) \mspace{-35mu}\prod_{\substack{\big(\hat{i},j,k,m\big)\\\in\mathcal{I}_n\times\mathcal{J}_n\times\mathcal{K}_n\times\mathcal{M}_m}}\mspace{-45mu}q^n_{V|U}\Big(\mathbf{v}\big(\hat{i},j,k,m\big)\Big|\mathbf{u}(\hat{i})\Big).\label{EQ:codebook_probability}
\end{equation}
The encoder and decoder are described next for any superposition codebook $\mathcal{B}_n\in\mathfrak{B}_n$.


\par\textbf{Encoder $\bm{f_n^{(\mathcal{B}_n)}}$:} The encoding function is based on the likelihood-encoder~\cite{Cuff_Song_Likelihood2016}, which allows us to approximate the induced joint distribution by a simple distribution that we use for the analysis. Given $m\in\mathcal{M}_n$ and $\mathbf{s}\in\mathcal{S}^n$, the encoder randomly chooses $(i,j,k)\in\mathcal{I}_n\times\mathcal{J}_n\times\mathcal{K}_n$ according to
\begin{equation}
	p_{\mathsf{LE}}^{(\mathcal{B}_n)}(i,j,k|m,\mathbf{s})=\frac{q^n_{S|U,V}\big(\mathbf{s}\big|\mathbf{u}(i),\mathbf{v}(i,j,k,m)\big)}{\mspace{-15mu}\sum\limits_{\substack{(i',j',k')\\\in\mathcal{I}_n\times\mathcal{J}_n\times\mathcal{K}_n}}\mspace{-15mu}q^n_{S|U,V}\big(\mathbf{s}\big|\mathbf{u}(i'),\mathbf{v}(i',j',k',m)\big)}, \label{EQ:main_proof_likelihood_enc}
\end{equation}
where $q_{S|U,V}$ is the conditional marginal of $q_{S,U,V}$ defined by $q_{S,U,V}(s,u,v)=\sum_{x\in\mathcal{X}}W_S(s)q_{U,V,X|S}(u,v,x|s)$, for every $(s,u,v)\in\mathcal{S}\times\mathcal{U}\times\mathcal{V}$. The encoder declares the chosen index $k\in\mathcal{K}_n$ as the key. The channel input sequence is generated by feeding the chosen $u$- and $v$-codewords along with the state sequence into the DM channel $q_{X|U,V,S}$, i.e., it is sampled from the random vector $\mathbf{X}\sim q^n_{X|U=\mathbf{u}(i),V=\mathbf{v}(i,j,k,m),S=\mathbf{s}}$.

Accordingly, the (stochastic) encoding function $f_n:\mathcal{M}_n\times\mathcal{S}^n\to\mathcal{P}(\mathcal{K}_n\times\mathcal{X}^n)$ is given by 
\begin{align*}
    f_n^{(\mathcal{B}_n)}(k,\mathbf{x}|m,\mathbf{s})=\mspace{-25mu}\sum_{(i,j)\in\mathcal{I}_n\times\mathcal{J}_n}\mspace{-10mu}&\Big[p_{\mathsf{LE}}^{(\mathcal{B}_n)}(i,j,k|m,\mathbf{s}) \numberthis\label{EQ:main_proof_encoder}\\
    \times & q_{X|U,V,S}^n\big(\mathbf{x}\big|\mathbf{u}(i),\mathbf{v}(i,j,k,m),\mathbf{s}\big)\Big].
\end{align*}


\par\textbf{Decoder $\bm{\phi_n^{(\mathcal{B}_n)}}$:} Upon observing $\mathbf{y}\in\mathcal{Y}^n$, the decoder searches for a unique tuple $(\hat{i},\hat{j},\hat{k},\hat{m})\in\mathcal{I}_n\times\mathcal{J}_n\times\mathcal{K}_n\times\mathcal{M}_n$ such that
\begin{equation}
    \Big(\mathbf{u}(\hat{i}),\mathbf{v}(\hat{i},\hat{j},\hat{k},\hat{m}),\mathbf{y}\Big)\in\mathcal{T}_\epsilon^{n}(q_{U,V,Y}).
\end{equation}
If such a unique quadruple is found, then set $\phi_n^{(\mathcal{B}_n)}(\mathbf{y})=\big(\hat{m},\hat{k}\big)$; otherwise, $\phi_n^{(\mathcal{B}_n)}(\mathbf{y})=(1,1)$.

The quadruple $(\mathcal{M}_n,\mathcal{K}_n,f_n^{(\mathcal{B}_n)},\phi_n^{(\mathcal{B}_n)})$ defined  with respect to the codebook $\mathcal{B}_n$ is an $(n,R_M,R_K)$-code $c_n$. For any message distribution $p_M\in\mathcal{P}(\mathcal{M}_n)$ and codebook $\mathcal{B}_n\in\mathfrak{B}_n$, the induced joint distribution $p^{(\mathcal{B}_n)}$ over $\mathcal{M}_n\times\mathcal{S}^n\times\mathcal{I}_n\times\mathcal{J}_n\times\mathcal{K}_n\times\mathcal{U}^n\times\mathcal{V}^n\times\mathcal{X}^n\times\mathcal{Y}^n\times\mathcal{Z}^n\times\hat{\mathcal{M}}_n\times\hat{\mathcal{K}}_n$ is
\begin{align*}
    & p^{(\mathcal{B}_n)}(m,\mathbf{s},i,j,k,\mathbf{u},\mathbf{v},\mathbf{x},\mathbf{y},\mathbf{z},\hat{m},\hat{k})\\
    &\quad =  p_M(m)W_S^n(\mathbf{s})p_{\mathsf{LE}}^{(\mathcal{B}_n)}(i,j,k|m,\mathbf{s})\\
    &\quad\qquad \times \mathds{1}_{\big\{\mathbf{u}=\mathbf{u}(i)\big\}\cap \big\{\mathbf{v}=\mathbf{v}(i,j,k,m)\big\}}  q^n_{X|U,V,S}(\mathbf{x}|\mathbf{u},\mathbf{v},\mathbf{s})\\
    &\quad\qquad \times W^n_{Y,Z|S,X}(\mathbf{y},\mathbf{z}|\mathbf{s},\mathbf{x})\mathds{1}_{\big\{\left(\hat{m},\hat{k}\right)=\phi_n^{(\mathcal{B}_n)}(\mathbf{y})\big\}}.\numberthis\label{EQ:main_proof_induced_PMF}
\end{align*}
If $p_M=p_{\mathcal{M}_n}^{(U)}$, i.e., the message distribution is uniform, we write $\bar{p}^{(\mathcal{B}_n)}$ instead of $p^{(\mathcal{B}_n)}$.
If $p^{(\mathcal{B}_n)}$ appears with no explicitly stated argument, it should be interpreted as $p^{(\mathcal{B}_n)}(m,\mathbf{s},i,j,k,\mathbf{u},\mathbf{v},\mathbf{x},\mathbf{y},\mathbf{z},\hat{m},\hat{k})$. This abbreviation is used for $\bar{p}^{(\mathcal{B}_n)}$ and the approximating distributions, stated next, as well.

\textbf{Approximating Distribution:} For each $p_M\in\mathcal{P}(\mathcal{M}_n)$ and $\mathcal{B}_n\in\mathfrak{B}_n$, define the distribution
\begin{align*}
&\pi^{(\mathcal{B}_n)}(m,i,j,k,\mathbf{u},\mathbf{v},\mathbf{s},\mathbf{x},\mathbf{y},\mathbf{z},\hat{m},\hat{k})\\
& \quad\triangleq p_M(m)\frac{1}{|\mathcal{I}_n||\mathcal{J}_n||\mathcal{K}_n|}\mathds{1}\mspace{-2mu}_{\big\{\mspace{-2mu}\mathbf{u}=\mathbf{u}(i)\mspace{-2mu},\mathbf{v}=\mathbf{v}(i,j,k,m)\big\}}\\
&\quad\quad\times q^n_{S|U,V}(\mathbf{s}|\mathbf{u},\mathbf{v})q^n_{X|U,V,S}(\mathbf{x}|\mathbf{u},\mathbf{v},\mathbf{s})\\
&\quad\quad\quad\quad\times W^n_{Y,Z|S,X}(\mathbf{y},\mathbf{z}|\mathbf{s},\mathbf{x})\mathds{1}_{\big\{\left(\hat{m},\hat{k}\right)=\phi_n^{(\mathcal{B}_n)}(\mathbf{y})\big\}}.\numberthis\label{EQ:main_proof_target_PMF}
\end{align*}
As before, $\bar{\pi}^{(\mathcal{B}_n)}$ stands for $\pi^{(\mathcal{B}_n)}$ when $p_M=p_{\mathcal{M}_n}^{(U)}$.
This distribution describes a setup where the codeword indices \((i,j,k)\) are chosen uniformly at random, whereas the state sequence \(\mathbf{s}\) is the output of a DM prefix channel \(q_{S|U,V}\). Consequently, the effective channel from $(U,V)$ to $(Y,Z)$ in the approximating setup is
\begin{align*}
&q_{Y,Z|U,V}(y,z|u,v) = \numberthis\label{EQ:main_proof_target_PMF_channel} \\
&\mspace{-7mu}\sum_{(s,x)\in \mathcal{S}\times \mathcal{X}} \mspace{-15mu}q_{S|U,V}(s|u,v) q_{X|U,V,S}(x|u,v,s)W_{Y,Z|S,X}(y,z|s,x).
\end{align*}
Notably, $q_{Y,Z|U,V}$ is not SD, which allows simple reliability and security analyses.
We subsequently show that for a random codebook $\mathsf{B}_n$ with appropriately chosen rates (see Lemma~\ref{LEMMA:good_approximation} below), $p^{(\mathsf{B}_n)}$ and $\pi^{(\mathsf{B}_n)}$ are close in total variation, with high probability. Therefore, one may analyze the code's performance with respect to either of the two. The simplicity of $\pi^{(\mathsf{B}_n)}$ makes it preferable for the analysis.

The following lemma states sufficient conditions for $\pi^{(\mathsf{B}_n)}$ to be a good approximation (in total variation) of $p^{(\mathsf{B}_n)}$ with double-exponential certainty.
\begin{lemma}[Sufficient Conditions for Approximation]\label{LEMMA:good_approximation}
If
\begin{subequations}
	\begin{align}
		R_1&>I(U;S),\label{EQ:main_proof_approx_rate_bound1}\\
		R_1+R_2 + R_K&>I(U,V;S),\label{EQ:main_proof_approx_rate_bound2}
	\end{align}\label{EQ:main_proof_approx_rate_bounds}
\end{subequations}
then there exist $\alpha _1,\alpha_2>0$, such that for any $n$ large enough
\begin{equation}
	\mathbb{P}_\mu\bigg(\max_{p_M\in\mathcal{P}(\mathcal{M}_n)}\Big|\Big|p^{(\mathsf{B}_n)}-\pi^{(\mathsf{B}_n)}\Big|\Big|_{\mathsf{TV}}> e^{-n\alpha_1}\bigg)\leq e^{- e^{n\alpha_2}}.\label{EQ:main_proof_approx_soft_covering}
\end{equation}
In particular, for any such $n$ it also holds that
\begin{equation}
\mathbb{E}_\mu\Big|\Big|\bar{p}^{(\mathsf{B}_n)}-\bar{\pi}^{(\mathsf{B}_n)}\Big|\Big|_{\mathsf{TV}}\leq e^{-n\alpha_1}+n\log\left(\frac{1}{\xi_S}\right)e^{-e^{n\alpha_2}},\label{EQ:main_proof_approx_soft_covering_expect}
\end{equation}
where $\xi_S=\min_{s\in\supp(W_S)}W_S(s)>0$. The subscript $\mu$ in $\mathbb{P}_\mu$ and $\mathbb{E}_\mu$ indicates that the probability measure and the expectation are taken with respect to the random codebook $\mathsf{B}_n\sim \mu$.
\end{lemma}

Lemma~\ref{LEMMA:good_approximation} essentially restates~\cite[Lemma 7]{Goldfeld_SDWTC2016} with the index $j$ therein replaced here with the pair $(j,k)$. The proof of Lemma~\ref{LEMMA:good_approximation} relies on the strong SCL for superposition codes and some basic properties of total variation. Due to the similarity to~\cite[Lemma 7]{Goldfeld_SDWTC2016} we omit the proof and the reader is referred to~\cite{Goldfeld_SDWTC2016}. 

Lemma~\ref{LEMMA:good_approximation} is key for analyzing the performance of the proposed code. The reliability analysis that is presented next exploits the convergence of the expected value from \eqref{EQ:main_proof_approx_soft_covering_expect} to show that the average error probability can be made arbitrarily small. The expurgation method~\cite[Theorem 7.7.1]{CovThom06} is used in a later stage of this proof to upgrade to a vanishing maximal error probability.


\textbf{Average Error Probability Analysis:} The average error probability\footnote{We slightly abuse notation here because $\bar{e}$ and $e_m$ are actually functions of the code $c_n$ rather than the codebook $\mathcal{B}_n$. However, since $\mathcal{B}_n$ uniquely defines $c_n$ we prefer this presentation for the sake of simplicity.} $\bar{e}(\mathcal{B}_n)$ associated with a codebook $\mathcal{B}_n$ is
\begin{align*}
\bar{e}(\mathcal{B}_n)&\triangleq \frac{1}{|\mathcal{M}_n|}\sum_{m\in\mathcal{M}_n}e_m(\mathcal{B}_n)\\
&   =\mathbb{P}_{\bar{p}^{(\mathcal{B}_n)}}\left(\left(\hat{M},\hat{K}\right)\neq (M,K)\right).\numberthis\label{EQ:main_proof_error_prob_P}
\end{align*}
Our next step is to establish that the expected value of $\bar{e}(\mathsf{B}_n)$ over the codebook ensemble is approximately the same under $\bar{p}$ and $\bar{\pi}$. Then, the expected average error probability under $\bar{\pi}$ is analyzed and shown to converge to zero as $n\to\infty$. Due to the simple structure of $\bar{\pi}$, this analysis requires nothing but standard typicality arguments. To do so we use the two following lemmas.

\begin{lemma}[Average Error Prob. Under $\bar{p}^{(\mathsf{B}_n)}$ and $\bar{\pi}^{(\mathsf{B}_n)}$]\label{LEMMA:error_prob_PvsGamma}
    The following relation holds:
    \begin{align*}
        &\Big|\mathbb{E}_\mu\mathbb{P}_{\bar{p}^{(\mathsf{B}_n)}}\left( {\scriptstyle (\hat{M},\hat{K}) \neq (M,K)}\right)
        -\mathbb{E}_\mu\mathbb{P}_{\bar{\pi}^{(\mathsf{B}_n)}}\left( {\scriptstyle (\hat{M},\hat{K}) \neq (M,K)} \right)\Big|\\
        &\mspace{200mu}\leq \mathbb{E}_\mu\big|\big|\bar{p}^{(\mathsf{B}_n)}-\bar{\pi}^{(\mathsf{B}_n)}\big|\big|_{\mathsf{TV}}.\numberthis\label{EQ:main_proof_TV_error_prob_sandwich_expect}
    \end{align*}
\end{lemma}

Lemma~\ref{LEMMA:error_prob_PvsGamma} is a simple consequence of the definition of total variation and the linearity of expectation. For the proof of Lemma~\ref{LEMMA:error_prob_PvsGamma} and the following Lemma~\ref{LEMMA:error_prob_Gamma}, the reader is referred to the \emph{Average Error Probability Analysis} part in Section VI-B of~\cite{Goldfeld_SDWTC2016}.
    
\begin{lemma}[Average Error Probability Under $\bar{\pi}^{(\mathsf{B}_n)}$]\label{LEMMA:error_prob_Gamma}
If the rate tuple $(R_M,R_K,R_1,R_2)$ satisfies
\begin{subequations}
\begin{align}
    R_M+R_K+R_2<I(V;Y|U),\label{EQ:main_proof_reliability_bound1}\\
    R_M+R_K+R_1+R_2<I(U,V;Y),\label{EQ:main_proof_reliability_bound2}
\end{align}\label{EQ:main_proof_reliability_bounds}
\end{subequations}
then
\begin{equation}
    \mathbb{E}_\mu\mathbb{P}_{\bar{\pi}^{(\mathsf{B}_n)}}\left(\left(\hat{M},\hat{K}\right)\neq (M,K)\right) \xrightarrow[n \to \infty]{} 0.
\end{equation} 
\end{lemma}

Since $\pi^{(\mathsf{B}_n)}$ describes a setup where the channel is not SD (see \eqref{EQ:main_proof_target_PMF}-\eqref{EQ:main_proof_target_PMF_channel}), standard typicality decoding arguments for superposition codes apply, and, in turn, imply the result of Lemma 3.
We stress that the conditions in \eqref{EQ:main_proof_reliability_bounds} ensure reliable decoding of the four indices $(i,j,k,m)$, and, in particular, of the SM-SK pair $(m,k)$. 

Combining the claims of Lemmas~\ref{LEMMA:error_prob_PvsGamma}-\ref{LEMMA:error_prob_Gamma} with  \eqref{EQ:main_proof_approx_soft_covering_expect} from Lemma~\ref{LEMMA:good_approximation}, we have that as long as \eqref{EQ:main_proof_reliability_bounds} and \eqref{EQ:main_proof_approx_rate_bounds} are satisfied
\begin{equation}
    \mathbb{E}_\mu \bar{e}(\mathsf{B}_n) \xrightarrow[n \to \infty]{} 0.
\end{equation}
    
\textbf{Key Analysis:}
The structure of $\pi^{(\mathcal{B}_n)}$ from  \eqref{EQ:main_proof_target_PMF} implies that for any $\mathcal{B}_n\in\mathfrak{B}_n$ and $m \in \mathcal{M}_n$ we have $\pi^{(\mathcal{B}_n)}_{K|M=m} = p^{(U)}_{\mathcal{K}_n}$. Adopting the same abuse of notation we used for the reliability analysis, we use Lemma~\ref{LEMMA:good_approximation} to upper bound the probability that $\delta(\mathsf{B}_n)$ does not decay exponentially fast to zero as $n$ grows. Therefore, assuming \eqref{EQ:main_proof_approx_rate_bounds} holds, we have that there exists $\eta_1,\eta_2>$ such that
\begin{align*}
    &\mathbb{P}_\mu\Big(\delta(\mathsf{B}_n)>e^{-n\eta_1}\Big)\\
    &\qquad=\mathbb{P}_\mu\Big(\max_{m\in\mathcal{M}_n}\Big|\Big|p^{(\mathsf{B}_n)}_{K|M=m} - p^{(U)}_{\mathcal{K}_n}\Big|\Big|_{\mathrm{TV}}>e^{-n\eta_1}\Big)\\
    &\qquad=\mathbb{P}_\mu\left(\max_{m\in\mathcal{M}_n}\Big|\Big|p^{(\mathsf{B}_n)}_{K|M=m} - \pi^{(\mathsf{B}_n)}_{K|M=m}\Big|\Big|_{\mathrm{TV}}>e^{-n\eta_1}\right)\\
    &\qquad\leq\mathbb{P}_\mu\left(\max_{p_M\in\mathcal{P}(\mathcal{M}_n)}\Big|\Big|p^{(\mathsf{B}_n)}_{M,K} - \pi^{(\mathsf{B}_n)}_{M,K}\Big|\Big|_{\mathrm{TV}}>e^{-n\eta_1}\right) \\
    &\qquad\leq\mathbb{P}_\mu\left(\max_{p_M\in\mathcal{P}(\mathcal{M}_n)}\Big|\Big|p^{(\mathsf{B}_n)} - \pi^{(\mathsf{B}_n)}\Big|\Big|_{\mathrm{TV}}>e^{-n\eta_1}\right) \\
    &\qquad\stackrel{(a)}\leq e^{-e^{n\eta_2}},\numberthis\label{EQ:key_analysis_faliure_probability}
\end{align*}
where (a) is by \eqref{EQ:main_proof_approx_soft_covering} from Lemma~\ref{LEMMA:good_approximation}. We proceed with the security analysis.
\ \\


\textbf{Security Analysis:} This part mainly deals with analyzing the SS metric under the distribution $\pi^{(\mathsf{B}_n)}$. The following lemma explains the reason for doing so. It states conditions under which SS under $\pi^{(\mathsf{B}_n)}$ implies SS under $p^{(\mathsf{B}_n)}$. These conditions are assured, with hight probability, by Lemma~\ref{LEMMA:good_approximation}.

\begin{lemma}[SS for $p^{(\mathcal{B}_n)}$ and $\pi^{(\mathcal{B}_n)}$]\label{LEMMA:ss_p_gamma}
    Let $\mathcal{B}_n\in\mathfrak{B}_n$ and $\beta_1>0$, such that for all $p_M\in\mathcal{P}(\mathcal{M}_n)$ and $n$ sufficiently large (independent of $p_M$)
    \begin{equation}
        \Big|\Big|p_M p^{(\mathcal{B}_n)}_{K,\mathbf{Z}|M} - p_M\pi^{(\mathcal{B}_n)}_{K,\mathbf{Z}|M}\Big|\Big|_{\mathrm{TV}}\leq e^{-n\beta_1}.\label{EQ:lemma_ss_if}
    \end{equation}
    Then, there exist $\beta_2>0$ such that for all $p_M\in\mathcal{P}(\mathcal{M}_n)$ and large enough values of $n$ (independent of $p_M$), we have
    \begin{equation} \Big|I_{p^{(\mathcal{B}_n)}}(M,K;\mathbf{Z})-I_{\pi^{(\mathcal{B}_n)}}(M,K;\mathbf{Z})\Big|\leq e^{-n\beta_2},\label{EQ:lemma_ss_then}
    \end{equation}
    where the subscripts $p^{(\mathcal{B}_n)}$ and $\pi^{(\mathcal{B}_n)}$ indicate that a mutual information term is calculated with respect to the corresponding PMF.
\end{lemma}
The proof of Lemma~\ref{LEMMA:ss_p_gamma} extends that of~\cite[Lemma 8]{Goldfeld_SDWTC2016}, and can be found in~\cite[Appendix D]{Bunin_SkSm_Arxiv}.

For any $n\in\mathbb{N}$ and $\beta_1>0$, define the collection of codebooks 
\begin{equation}
	\mathcal{A}_n\left(\beta_1\right) \triangleq \hspace{-3mm}
	 \quad {\left\{\mathcal{B}_n\bigg| \max_{p_M\in\mathcal{P}(\mathcal{M}_n)}\Big|\Big|p^{(\mathcal{B}_n)} - \pi^{(\mathcal{B}_n)}\Big|\Big|_{\mathrm{TV}} \le e^{-n\beta_1}	\right\}}.
\end{equation}
We note that Lemma~\ref{LEMMA:good_approximation} guarantees that if \eqref{EQ:main_proof_approx_rate_bounds} is satisfied, then there exist \(\beta_1>0\) such that \(\mathbb{P}_\mu\Big(\mathsf{B}_n\notin\mathcal{A}_n\left(\beta_1\right)\Big)\) vanishes doubly exponentially fast with $n$.
Lemma~\ref{LEMMA:ss_p_gamma} then ensures that if $\mathcal{B}_n\in\mathcal{A}_n(\beta_1)$, for some $\beta_1>0$ and sufficiently large $n$, then there exists $\beta_2>0$, such that
\begin{align*}
    \ell_\mathsf{Sem} (\mathcal{B}_n) &\triangleq \max_{p_M\in\mathcal{P}(\mathcal{M}_n)}I_{p^{(\mathcal{B}_n)}}(M,K;\mathbf{Z})\\
    &\leq \max_{p_M\in\mathcal{P}(\mathcal{M}_n)}I_{\pi^{(\mathcal{B}_n)}}(M,K;\mathbf{Z})+e^{-n\beta_2},\numberthis\label{EQ:main_proof_SSP_UB2}
\end{align*}
for large enough $n$. Therefore, to demonstrate that the code corresponding to any  $\mathcal{B}_n\in\mathcal{A}_n(\beta_1)$ is semantically-secured it suffices to show that $\max_{p_M\in\mathcal{P}(\mathcal{M}_n)}I_{\pi^{(\mathcal{B}_n)}}(M,K;\mathbf{Z})$ can be made arbitrarily small.

Fix $\mathcal{B}_n\in\mathcal{A}_n(\beta_1)$ and $p_M\in\mathcal{P}(\mathcal{M}_n)$, and consider
\begin{align*}
    I_{\pi^{(\mathcal{B}_n)}}&(M,K;\mathbf{Z})\leq I_{\pi^{(\mathcal{B}_n)}}(M,K;I,\mathbf{U},\mathbf{Z})\\
                          &=\mathsf{D}\Big(\pi^{(\mathcal{B}_n)}_{M,K,\mathbf{Z},I,\mathbf{U}}\Big|\Big|\pi^{(\mathcal{B}_n)}_{M,K}\pi^{(\mathcal{B}_n)}_{\mathbf{Z},I,\mathbf{U}}\Big)\\
                          &\stackrel{(a)}=\mathsf{D}\Big(\pi^{(\mathcal{B}_n)}_{M,K}\pi^{(\mathcal{B}_n)}_{I,\mathbf{U}}\pi^{(\mathcal{B}_n)}_{\mathbf{Z}|M,K,I,\mathbf{U}}\Big|\Big|\pi^{(\mathcal{B}_n)}_{M,K}\pi^{(\mathcal{B}_n)}_{I,\mathbf{U}}\pi^{(\mathcal{B}_n)}_{\mathbf{Z}|I,\mathbf{U}}\Big)\\
                          &\stackrel{(b)}=\mathsf{D}\Big(\pi^{(\mathcal{B}_n)}_{\mathbf{Z}|M,K,I,\mathbf{U}}\Big|\Big|\pi^{(\mathcal{B}_n)}_{\mathbf{Z}|I,\mathbf{U}}\Big|\pi^{(\mathcal{B}_n)}_{M,K}\pi^{(\mathcal{B}_n)}_{I,\mathbf{U}}\Big)\\
                          &\stackrel{(c)}\leq \mathsf{D}\Big(\pi^{(\mathcal{B}_n)}_{\mathbf{Z}|M,K,I,\mathbf{U}}\Big|\Big|q^n_{Z|U}\Big|\pi^{(\mathcal{B}_n)}_{M,K}\pi^{(\mathcal{B}_n)}_{I,\mathbf{U}}\Big),\numberthis\label{EQ:main_proof_SS_gamma_MI_UB}
\end{align*}
where (a) is because $\pi^{(\mathcal{B}_n)}_{M,K,I,\mathbf{U}}=\pi^{(\mathcal{B}_n)}_{M,K}\pi^{(\mathcal{B}_n)}_{I,\mathbf{U}}$ (see \eqref{EQ:main_proof_target_PMF}), (b) is by the relative entropy chain rule, while (c) follows from
\begin{align*}
    &\mathsf{D}\Big(\pi^{(\mathcal{B}_n)}_{\mathbf{Z}|M,K,I,\mathbf{U}}\Big|\Big|\pi^{(\mathcal{B}_n)}_{\mathbf{Z}|I,\mathbf{U}}\Big|\pi^{(\mathcal{B}_n)}_{M,K}\pi^{(\mathcal{B}_n)}_{I,\mathbf{U}}\Big)\\
    &\qquad\qquad=\mathsf{D}\Big(\pi^{(\mathcal{B}_n)}_{\mathbf{Z}|M,K,I,\mathbf{U}}\Big|\Big|q^n_{Z|U}\Big|\pi^{(\mathcal{B}_n)}_{M,K}\pi^{(\mathcal{B}_n)}_{I,\mathbf{U}}\Big)\\
        &\qquad\qquad\qquad-\mathsf{D}\Big(\pi^{(\mathcal{B}_n)}_{\mathbf{Z}|I,\mathbf{U}}\Big|\Big|q^n_{Z|U}\Big|\pi^{(\mathcal{B}_n)}_{M,K}\pi^{(\mathcal{B}_n)}_{I,\mathbf{U}}\Big)
    \numberthis\label{EQ:main_proof_relative_entropy_q}
\end{align*}
and the non-negativity of relative entropy. Here,  $q_{Z|U}$ is the  conditional marginal of the single-letter distribution $W_Sq_{U,V,X|S}W_{Y,Z|S,X}$.
    
Maximizing both sides of \eqref{EQ:main_proof_SS_gamma_MI_UB} over all message distributions $p_M\in\mathcal{P}(\mathcal{M}_n)$, we further have
\begin{align*}
    &\max_{p_M\in\mathcal{P}(\mathcal{M}_n)}I_\pi(M,K;\mathbf{Z})\\
    &\leq\max_{p_M\in\mathcal{P}(\mathcal{M}_n)}\mathsf{D}\Big(\pi^{(\mathcal{B}_n)}_{\mathbf{Z}|M,K,I,\mathbf{U}}\Big|\Big|q^n_{Z|U}\Big|\pi^{(\mathcal{B}_n)}_{M,K}\pi^{(\mathcal{B}_n)}_{I,\mathbf{U}}\Big)\\
    &= \max_{p_M\in\mathcal{P}(\mathcal{M}_n)}\mspace{-10mu}\sum_{(m,k)\in\mathcal{M}_n\times\mathcal{K}_n}\Bigg[\pi^{(\mathcal{B}_n)}_{M,K}(m,k)\\
    &\qquad\qquad\times\mathsf{D}\Big(\pi^{(\mathcal{B}_n)}_{\mathbf{Z}|M=m,K=k,I,\mathbf{U}}\Big|\Big|q^n_{Z|U}\Big|\pi^{(\mathcal{B}_n)}_{I,\mathbf{U}}\Big)\Bigg]\\
    &\le \max_{p_M\in\mathcal{P}(\mathcal{M}_n)}\mspace{-10mu}\sum_{(m,k)\in\mathcal{M}_n\times\mathcal{K}_n}\Bigg[\pi^{(\mathcal{B}_n)}_{M,K}(m,k)\\
    &\qquad\qquad\times\mspace{-15mu}\max_{(\tilde{m},\tilde{k})\in\mathcal{M}\times\mathcal{K}_n}
    \mathsf{D}\Big(\pi^{(\mathcal{B}_n)}_{\mathbf{Z}|M=\tilde{m},K=\tilde{k},I,\mathbf{U}}\Big|\Big|q^n_{Z|U}\Big|\pi^{(\mathcal{B}_n)}_{I,\mathbf{U}}\Big)\Bigg] \\
    &=\max_{(m,k)\in\mathcal{M}_n\times\mathcal{K}_n}\mathsf{D}\Big(\pi^{(\mathcal{B}_n)}_{\mathbf{Z}|M=m,K=k,I,\mathbf{U}}\Big|\Big|q^n_{Z|U}\Big|\pi^{(\mathcal{B}_n)}_{I,\mathbf{U}}\Big). \numberthis\label{EQ:main_proof_relative_ent_max_m}
\end{align*}
		Inserting \eqref{EQ:main_proof_relative_ent_max_m} into \eqref{EQ:main_proof_SSP_UB2}, for a sufficiently large $n$, we deduce there exists $\beta_2>0$ such that
\begin{align*}
    &\ell_\mathsf{Sem} (\mathcal{B}_n) \le \numberthis\label{EQ:main_proof_bounding_ss_with_realtive_entropy}\\
    &\max_{(m,k)\in\mathcal{M}_n\times\mathcal{K}_n}\mathsf{D}\Big(\pi^{(\mathcal{B}_n)}_{\mathbf{Z}|M=m,K=k,I,\mathbf{U}}\Big|\Big|q^n_{Z|U}\Big|\pi^{(\mathcal{B}_n)}_{I,\mathbf{U}}\Big) + e^{-n\beta_2}.
\end{align*}

The two following lemmas state conditions under which the probability that the RHS of \eqref{EQ:main_proof_bounding_ss_with_realtive_entropy} vanishes exponentially fast with $n$ is double-exponentially close to~1.
    
\begin{lemma}[Total Variation Dominates Relative Entropy]\label{LEMMA:TV_exp_decay}
    Let $\mathcal{X}$ and $\mathcal{Y}$ be finite sets, and for any $n\in\mathbb{N}$ let $p_\mathbf{X}\in\mathcal{P}(\mathcal{X}^n)$, $p_{\mathbf{Y}|\mathbf{X}}:\mathcal{X}^n\to\mathcal{P}(\mathcal{Y}^n)$ and $q_{Y|X}:\mathcal{X}\to\mathcal{P}(\mathcal{Y})$. If $p_{\mathbf{Y}|\mathbf{X}=\mathbf{x}}\ll q^n_{Y|X=\mathbf{x}}$, for all $\mathbf{x}\in\mathcal{X}^n$, i.e., $p_{\mathbf{Y}|\mathbf{X}=\mathbf{x}}$ is absolutely continuous with respect to $q^n_{Y|X=\mathbf{x}}$, then
    \begin{align*}
        &\mathsf{D}\big(p_{\mathbf{Y}|\mathbf{X}}\big|\big|q^n_{Y|X}\big|p_\mathbf{X}\big)\leq\big|\big|p_\mathbf{X}p_{\mathbf{Y}|\mathbf{X}}-p_\mathbf{X}q^n_{Y|X}\big|\big|_{\mathrm{TV}} \numberthis\\
        &\ \times \mspace{-7mu} \left(\mspace{-4mu} n\log|\mathcal{Y}|+\log\frac{1}{\big|\big|p_\mathbf{X}p_{\mathbf{Y}|\mathbf{X}}-p_\mathbf{X}q^n_{Y|X}\big|\big|_{\mathrm{TV}}}+n\log\xi_{Y|X}\mspace{-4mu}\right),
    \end{align*}
    where $\xi_{Y|X}$ is the minimal non-zero value of the transition matrix $q_{Y|X}$.
\end{lemma}
Lemma~\ref{LEMMA:TV_exp_decay} is~\cite[Lemma 9]{Goldfeld_SDWTC2016} and its proof is omitted.

It is readily verified that
$\pi^{(\mathcal{B}_n)}_{\mathbf{Z}|M=m,K=k,I=i,\mathbf{U}=\mathbf{u}}\ll q^n_{Z|U=\mathbf{u}}$, for each $(m,i,k,\mathbf{u})\in\mathcal{M}_n\times\mathcal{I}_n\times\mathcal{K}_n\times\mathcal{U}^n$. Combining Lemma~\ref{LEMMA:TV_exp_decay} and \eqref{EQ:main_proof_bounding_ss_with_realtive_entropy}, we see that if $\mathcal{B}_n\in\mathcal{A}_n(\beta_1)$ and
\begin{subequations} \label{EQ:main_proof_bound_lSEM_with_maxD}
\begin{equation}
    \max\limits_{\substack{(m,k)\\\in\mathcal{M}_n\times\mathcal{K}_n}}\Big|\Big| \pi^{(\mathcal{B}_n)}_{I,\mathbf{U}}\pi^{(\mathcal{B}_n)}_{\mathbf{Z}|M=m,K=k,I,\mathbf{U}}-\pi^{(\mathcal{B}_n)}_{I,\mathbf{U}}q^n_{Z|U}\Big|\Big|_{\mathrm{TV}} \le e^{-n\zeta_1},
\end{equation}
for some $\beta_1,\zeta_1>0$ and $n$ sufficiently large, then there exists $\zeta_2>0$ for which
\begin{equation}
    \ell_\mathsf{Sem} (\mathcal{B}_n) \le e^{-n\zeta_2}
\end{equation}\end{subequations} 
as $n$ grows.

\begin{lemma}[Sufficient Conditions for SS]\label{LEMMA:sufficient_condtions_ss}
    If the rate tuple $(R_M,R_K,R_1,R_2)\in\mathbb{R}^4_+$ satisfies \eqref{EQ:main_proof_approx_rate_bound1} and
    \begin{equation}
        R_2 > I(V;Z|U), \label{EQ:main_proof_SS_RB_final}
    \end{equation}
    then there exist $\gamma_1,\gamma_2>0$, such that for $n$ sufficiently large
     \begin{align*}
        \mathbb{P}_\mu\Bigg(\max\limits_{\substack{(m,k)\\\in\mathcal{M}_n\times\mathcal{K}_n}}\mspace{-8mu}\Big|\Big| &\pi^{(\mathsf{B}_n)}_{I,\mathbf{U}}\pi^{(\mathsf{B}_n)}_{\mathbf{Z}|M=m,K=k,I,\mathbf{U}}\\
        &-\pi^{(\mathsf{B}_n)}_{I,\mathbf{U}}q^n_{Z|U}\Big|\Big|_{\mathrm{TV}}>e^{-n\gamma_1}\Bigg)
        \leq e^{-e^{n\gamma_2}}.\numberthis\label{EQ:main_proof_gamma_SS_done}
    \end{align*}   
\end{lemma}
Lemma~\ref{LEMMA:sufficient_condtions_ss} follows by the security analysis from~\cite{Goldfeld_SDWTC2016} with $(M,K)=(m,k)$ in the role of $M=m$ therein.

Combining the lemma with Lemma~\ref{LEMMA:good_approximation} and \eqref{EQ:main_proof_bound_lSEM_with_maxD}, we deduce that if \eqref{EQ:main_proof_approx_rate_bounds} and \eqref{EQ:main_proof_SS_RB_final} hold, then there exist $\tau_1,\tau_2,\tau_3,\tau_4,\tau_5>0$ (dependent among themselves but independent of $n$), such that for any sufficiently large $n$
\begin{align*}
    &\mathbb{P}_\mu\Big(\ell_\mathsf{Sem} (\mathsf{B}_n) > e^{-n\tau_1}\Big)\\
    &\le\mathbb{P}_\mu\mspace{-1mu}\Big(\ell_\mathsf{Sem} (\mathsf{B}_n)\mspace{-3mu}>\mspace{-3mu}e^{-n\tau_1} \Big| \mathsf{B}_n\mspace{-3mu}\in\mspace{-3mu}\mathcal{A}_n\mspace{-3mu}\left(\tau_3\right)\mspace{-3mu}\Big)\mspace{-2mu}+\mspace{-1mu}\mathbb{P}_\mu\Big(\mspace{-2mu}\mathsf{B}_n\mspace{-2mu}\notin\mspace{-2mu}
    \mathcal{A}_n\mspace{-3mu}\left(\tau_3\right)\mspace{-4mu}\Big)\\
    &\le e^{-e^{n\tau_4}} + e^{-e^{n\tau_5}} \le e^{-e^{n\tau_2}}.\numberthis
\end{align*}
        
\ \\
\textbf{Code Extraction:} The above derivation shows that if \eqref{EQ:main_proof_approx_rate_bounds}, \eqref{EQ:main_proof_reliability_bounds} and \eqref{EQ:main_proof_SS_RB_final} are simultaneously satisfied, then
\begin{subequations}
\begin{equation}
    \mathbb{E}_{\mu}\bar{e}(\mathsf{B}_n)\xrightarrow[n\to\infty]{}0,\label{EQ:average_error_prob_final}
\end{equation}
and for sufficiently large $n$, we also have
\begin{align}
    \mathbb{P}_\mu\Big(\delta(\mathsf{B}_n)&>e^{-n\eta_1}\Big) \leq e^{-e^{n\eta_2}}, \label{EQ:main_proof_delta}\\        \mathbb{P}_\mu\Big(\ell_\mathsf{Sem}(\mathsf{B}_n)&>e^{-n\tau_1}\Big) \leq e^{-e^{n\tau_2}} \label{EQ:main_proof_lSEM}.
\end{align}\label{EQ:main_proof_results_sum}
\end{subequations}

The Selection Lemma from\cite[Lemma 5]{Goldfeld_WTCII2016} implies the existence of a sequence of superposition codebooks $\big\{\mathcal{B}_n\big\}_{n\in\mathbb{N}}$ (an outcome of the random codebook sequence $\big\{\mathsf{B}_n\big\}_{n\in\mathbb{N}}$), for which
\begin{subequations}
\begin{align}
    \bar{e}(\mathcal{B}_n)&\xrightarrow[n\to\infty]{}0,\label{EQ:main_proof_average_error_prob_vanish_fixed_code}\\
    \mathds{1}_{\big\{\delta(\mathcal{B}_n)>e^{-n\eta_1}\big\}}&\xrightarrow[n\to\infty]{}0,\label{EQ:main_proof_delta_fixed_code}\\
    \mathds{1}_{\big\{\ell_\mathsf{Sem}(\mathcal{B}_n)>e^{-n\tau_1}\big\}}&\xrightarrow[n\to\infty]{}0.\label{EQ:main_proof_lSEM_fixed_code}
\end{align}\label{EQ:main_proof_results_fixed_code}
\end{subequations}
Since the indicator functions in \eqref{EQ:main_proof_delta_fixed_code}-\eqref{EQ:main_proof_lSEM_fixed_code} take only the values 0 and 1, we have that for any $n$ large enough
\begin{subequations}
\begin{align}
    \delta(\mathcal{B}_n)\leq e^{-n\eta_1},\label{EQ:key_final}\\
    \ell_\mathsf{Sem}(\mathcal{B}_n)\leq e^{-n\tau_1}.\label{EQ:semantic_final}
\end{align}\label{EQ:key_semantic_final}
\end{subequations}
On account of \eqref{EQ:average_error_prob_final} and \eqref{EQ:key_semantic_final}, we have that $\{\mathcal{B}_n\}_{n\in\mathbb{N}}$ is semantically-secured, satisfies the target key statistics, and is reliable with respect to the \emph{average} error probability.
    
Our last step is to upgrade $\{\mathcal{B}_n\}_{n\in\mathbb{N}}$ to have a small \emph{maximal} error probability. This is a standard step that uses the expurgation technique (see, e.g.,~\cite[Theorem 7.7.1]{CovThom06}). Namely, pushing the average error probability below $\frac{\epsilon}{2}$, at least half of the messages in $\mathcal{M}_n$ result in a probability of error that is at most $\epsilon$. Throwing away the rest of the messages ensures a maximal error probability that is at most $\epsilon$, while inflicting a negligible rate loss. Discarding those messages does not harm the SS or the key uniformity and independence metric, thus producing a new sequence of codes that satisfies \eqref{EQ:achievability_def}. Applying the Fourier-Motzkin Elimination on \eqref{EQ:main_proof_approx_rate_bounds}, \eqref{EQ:main_proof_reliability_bounds} and \eqref{EQ:main_proof_SS_RB_final} shows that any SM-SK rate pair $(R_M,R_K)\in \mathcal{R}_\mathsf{A}\left(q_{U,V,X|S}\right)$ is achievable, which concludes the proof.


\section{Summary and Concluding Remarks}\label{SEC:summary}
	
We studied the trade-off between the SM and SK rates that are simultaneously achievable over a SD-WTC with non-causal encoder CSI. This model subsumes all other instances of CSI availability as special cases. An inner bound on the SS SM-SK capacity region was derived based on a superposition coding scheme, the likelihood encoder and soft-covering arguments inspired by~\cite{Goldfeld_SDWTC2016}.

We presented a class of SD-WTCs for which our inner bound achieves capacity, and showed that for this class, the previously best known SM-SK trade-off region by Prabhakaran \textit{et al.}~\cite{Prabhakaran_SKSM2012} is strictly sub-optimal. Furthermore, we showed that the inner bound derived here recovers the best lower bounds on either the SM~\cite{Goldfeld_SDWTC2016} or the SK~\cite{Bassi2016secretKey} rate achievable over the considered SD-WTC. Our derivations ensure SS, thus upgrading the security standard from most of the past results, which were derived under the weak secrecy metric.

As the SM-SK capacity region for this setup remains an open problem, good outer bounds are of particular interest. Extensions to multiple terminals, action dependent states~\cite{Han_Vinck_SDWTC_Actions2013}, and source reconstruction models should be examined as well.

\appendices

	\section{Proof of Corollary~\ref{corol:EveHasBetterChannel_SK}} \label{APPEN:proof_corol_EveHasBetterChannel_SK}
	
    Recall that the SD less-noisy-eavesdropper WTC with a key is the $\left(\tilde{\mathcal{S}},\mathcal{X},\tilde{\mathcal{Y}},\mathcal{Z}, W_{\tilde{S}}, W_{\tilde{Y},Z|\tilde{S},X}\right)$ SD-WTC, where $\tilde{\mathcal{S}}=\mathcal{L}\times\mathcal{S}$, $\tilde{\mathcal{Y}}=\mathcal{L}\times\mathcal{Y}$,
	$W_{\tilde{S}}=W_L\times W_S$, $\tilde{S}=(L,S)$, $\tilde{Y}=(L',Y)$, whose transition matrix satisfies \eqref{EQ:SD_LNE_WTC_WK_channel} and the less-noisy condition. 
    
    A \(q_{U,X|S,L}\) induces a joint distribution over $\mathcal{L}\times\mathcal{S}\times\mathcal{U}\times\mathcal{X}\times\mathcal{Y}\times\mathcal{Z}$ that is given by
        \begin{equation}
            q_{L,S,U,X,Y,Z} \triangleq W_L W_S q_{U,X|S,L} W_{Y,Z|S,X}. \label{EQ:EveHasBetterChannel_SK_jointPMF}
        \end{equation}
    We now proceed with the direct and the converse proofs.
    
    \textbf{Direct:}
    Fix \(q_{U,X|S}\) such that $(U,X)\mkv S \mkv L$. The structure of \eqref{EQ:EveHasBetterChannel_SK_jointPMF} further implies that $(S,U,X,Y,Z) \perp L$. Evaluating the bounds from Theorem~\ref{TM:SDWTC_lower_bound} with respect to \eqref{EQ:EveHasBetterChannel_SK_jointPMF}, while setting $V = (L,U)$ and using $\tilde{S}=(L,S)$ and $\tilde{Y}=(L,Y)$, we have
   \begin{subequations}
    \begin{align*}
        R_M &\le I(U,V;\tilde{Y}) - I(U,V;\tilde{S})\\
            &= I(L,U;L,Y) - I(L,U;L,S)\\
            &= I(U;Y|L) - I(U;S|L)\\
            &\stackrel{(a)}= I(U;Y) - I(U;S),\numberthis\label{EQ:SD_LNE_WTC_direct1}
    \end{align*}
    where (a) is because $(S,U,Y)$ are independent of $L$. Combining the two bounds on the sum $R_M+R_K$ in one, we further have
    \begin{align*}
        &R_K\hspace{-0.6mm}+\hspace{-0.5mm}R_M \le I(V;\tilde{Y}|U) \hspace{-0.6mm}-\hspace{-0.5mm} I(V;Z|U) \hspace{-0.5mm}-\hspace{-0.5mm} \big[ I(U;\tilde{S}) \hspace{-0.6mm}-\hspace{-0.5mm} I(U;\tilde{Y}) \big]^{+}\\
            & \ = I(L;L,Y|U) - I(L;Z|U) - \big[ I(U;L,S) - I(U;L,Y) \big]^{+}\\
            & \ \stackrel{(a)}= H(L) - \big[ I(U;S) - I(U;Y) \big]^{+},\numberthis\label{EQ:SD_LNE_WTC_direct2}
    \end{align*}
    \end{subequations}
    where, similarly to the above, (a) is implied by the independence of $(S,U,Y,Z)$ and $L$.
    Finally, due to \eqref{EQ:SD_LNE_WTC_direct1}, any joint distribution that produces a non-zero achievable region satisfies $I(U;Y) - I(U;S) \ge 0$; hence, the term $\big[ I(U;S) - I(U;Y) \big]^{+}$ from \eqref{EQ:SD_LNE_WTC_direct2} is zero. Maximizing over all \(q_{U,X|S}\) concludes the proof.

    \textbf{Converse:} To get \eqref{EQ:EveHasBetterChannel_SK_boundReliability}, notice that the secret communication rate of the setup cannot exceed the total reliable communication rate. Therefore, an upper bound on the SM capacity is given by the GP channel capacity formula~\cite{Gelfand_Pinsker}: \begin{equation}
        \max_{q_{U,X|\tilde{S}}} \Big[I(U;\tilde{Y}) - I(U;\tilde{S})\Big],
    \end{equation}
    where, for each $q_{U,X|\tilde{S}}$, the underlying joint PMF is $q_{U,X|\tilde{S}}W_{\tilde{Y}|\tilde{S},X}$, with $\tilde{S}=(L,S)$ and $\tilde{Y}=(L,Y)$. We thus have
    \begin{align*}
        R_M&\leq \max_{q_{U,X|L,S}} \big[I(U;L,Y) - I(U;L,S)\big]\\
        &= \max_{q_{U,X|L,S}} \big[I(U;Y|L) - I(U;S|L)\big]\\                
        &\stackrel{(a)}=\max_{q_{U,X|L,S}} \big[I(U;Y|L) - I(L,U;S)\big]\\
        &\le \max_{q_{U,X|L,S}} \big[I(L,U;Y) - I(L,U;S)\big]\\
        &\le \max_{q_{L,U,X|S}} \big[I(L,U;Y) - I(L,U;S)\big]\\
        &\stackrel{(b)}= \max_{q_{U,X|S}} \big[I(U;Y) - I(U;S)\big],\numberthis \label{EQ:EveHasBetterChannel_SK_GP1}         
    \end{align*}
    where (a) follows because $L$ and $S$ are independent (see \eqref{EQ:EveHasBetterChannel_SK_jointPMF}), while (b) follows by recasting $(L,U)$ as $U$.
    
    For the bound on $R_M+R_K$ from \eqref{EQ:EveHasBetterChannel_SK_boundSecrecy}, consider
    \begin{align*}
		&H(M,K) \\ 
		&\stackrel{(a)}\le I(M,K;\mathbf{L},\mathbf{Y}) + H(M,K|\mathbf{L},\mathbf{Y}) - I(M,K;\mathbf{Z}) +n\tilde{\epsilon}_n \\
		&\stackrel{(b)}\le I(M,K;\mathbf{L},\mathbf{Y}) - I(M,K;\mathbf{Z}) +n\epsilon_n\\
		&= I(M,K;\mathbf{L}|\mathbf{Y}) + I(M,K;\mathbf{Y}) - I(M,K;\mathbf{Z}) +n\epsilon_n\\
		&\stackrel{(c)}\le I(M,K;\mathbf{L}|\mathbf{Y}) + n\epsilon_n \le n(H(L)+ \epsilon_n), \numberthis
		\end{align*}
		where (a) uses the security hypothesis; (b) is Fano's inequality; whereas (c) follows the \emph{less-noisy} property of the channel since \((M,K) \mkv \mathbf{X} \mkv (\mathbf{Y},\mathbf{Z})\) is a Markov chain.
		
    Finally, since the code guarantees reliable communication for any message distribution, we can consider the case that it is uniform, while the key distribution (approximate) uniformity is guaranteed by the key properties. Thus
    \begin{equation}
    	R_M + R_K \le \frac{1}{n}H(M,K)+\hat{\epsilon}_n \le H(L) + \hat{\hat{\epsilon}}_n,
    \end{equation}
    which concludes the proof.

	\section{Counterexample to Theorem 1 from~\cite{Zibaeenejad2015WtSiPublicComm}} \label{APPEN:Zib}
	
	We first restate~\cite[Theorem 1]{Zibaeenejad2015WtSiPublicComm} through the notations of this work. This theorem proposes the following lower bound on the SK capacity $C^{\mathsf{SK}}$ of the SD-WTC with non-causal encoder CSI:
	\footnote{
		\cite[Theorem 1]{Zibaeenejad2015WtSiPublicComm} considers a setting with state observations at the receiver and the eavesdropper, and a public communication link. As explained in Remark~\ref{REM:most_general_setting}, such a setup is a special case of the GP-WTC. Using the technique described in Remark~\ref{REM:zib_main}, it can be verified that~\cite[Theorem 1]{Zibaeenejad2015WtSiPublicComm} (in its original form) is recoverable from its restatement here.
    }
	\begin{subequations} \label{EQ:Zib}
	    \begin{equation}
	        C^{\mathsf{SK}} \ge R_\mathsf{Zib} \triangleq \max \Big[ I(V;Y|U)-I(V;Z|U)\Big],
	    \end{equation}
	    where the maximization is over all conditional PMFs $q_{U|V}:\mathcal{V}\to\mathcal{P}(\mathcal{U})$ and $q_{V,X|S}:\mathcal{S}\to\mathcal{P}(\mathcal{V}\times\mathcal{X})$ satisfying
	    \begin{equation}
	        I(V;Y) \ge I(V;S).
	    \end{equation}
		All the above MI terms are taken with respect to the appropriate marginals of $W_Sq_{U|V}q_{V,X|S}W_{Y,Z|S,X}$, where $U \mkv V \mkv (S,X) \mkv (Y,Z)$ forms a Markov chain.
	\end{subequations}
	
		We next show that \eqref{EQ:Zib} cannot be an inner bound on the SK capacity of the GP-WTC. This is proven by constructing an example for which $R_{\mathsf{Zib}}$ exceeds the SK capacity. Consider the following:
	\begin{itemize}
	    \item Let $A$, $B$ and $Q$ be three i.i.d. $\mathsf{Ber}(\frac{1}{2})$ random variables. Also, set  $A^n$, $B^n$ and $Q^n$ as three $n$-fold random vectors whose coordinates are i.i.d. copies of $A$, $B$ and $Q$, respectively.
    	\item For each $i\in[1:n]$, let $T_i=t(A_i,B_i,Q_i)$,
    	        where $t:\{0,1\}^3 \to \{0,1\}$ is the deterministic function
    	        \begin{equation}
    	        t(a,b,q) = \begin{cases}
    						a ,\quad q=0  \\
    						b ,\quad q=1
    					\end{cases}.
    	        \end{equation}
	    \item Let $f_n$ be the stochastic encoder and $\Psi^n$ be the binary sequence that $f_n$ produces and transmits over a private binary bit-pipe to the legitimate receiver.
	    \item The encoder observes \((A^n,B^n)\) non-causally and determines the binary bit-pipe transmission \(\Psi^n\).
	    \item The decoder observes \((Q^n,T^n,\Psi^n)\).
	    \item The eavesdropper observes \(A^n \oplus_n B^n\), where $\oplus_n$ stands for bit-wise addition modulo 2. (At each time instance the eavesdropper observes $A_i + B_i \text{ (mod 2)}$.)
	\end{itemize}
	Thus, at each channel use $i\in[1:n]$, the encoder observes two fair coin tosses, $A_i$ and $B_i$. The decoder observes only one of them, namely $T_i$, chosen at random (using a third fair coin $Q_i$). The decoder knows which coin it observes, but the encoder does not. There is a \emph{private} bit-pipe from the encoder to the decoder, which enables the transmission of a single noiseless bit each time the coins are flipped.
	The legitimate parties wish to agree upon a key that is kept secret from the eavesdropper, who observes only the modulo 2 addition of the two coins, $A_i\oplus B_i$, each time they are flipped.
	
	Denoting the SK generated by the legitimate parties by $K_n$, the induced joint PMF of the system is
	\begin{align*}
	    &q_{A^n,B^n,Q^n,T^n,\Psi^n,K_n}(a^n,b^n,q^n,t^n,\psi^n,k_n) \\
	    &\ \ = f_n(k_n,\psi^n|a^n,b^n) \\
	    &\qquad \times \prod_{i=1}^n \Big[W_A(a_i)W_B(b_i)W_Q(q_i)\mathds{1}_{\{T_i=t(a_i,b_i,q_i)\}}  \Big]. \numberthis \label{EQ:Zib_joint_PMF}
	\end{align*}
	To see that the example falls within the framework of our model, note that \((A,B,T,Q)\) are correlated random sources (i.i.d. across time), such that the encoder, decoder and eavesdropper observe $(A,B)$, $(T,Q)$ and $A \oplus B$, respectively.
	In addition, there is a noiseless channel, independent of the sources, between the legitimate parties.
	In the notation of Remark~\ref{REM:most_general_setting} this corresponds to $S_t=(A,B)$, $S_r=(T,Q)$, $S_e=A \oplus B$, $X=\tilde{Y}=\Psi$ and $\tilde{Z}=0$ , such that:
	\begin{align*}
	W_{S_r,S_e|S_t} \mspace{-2.5mu}=\mspace{-2.5mu} W_{(Q,T),S_e|A,B} \mspace{-1.5mu}&=\mspace{-1.5mu} W_Q \mathds{1}_{\{T=t(A,B,Q)\}} \mathds{1}_{\{S_e = A \oplus B\}},\\
	W_{\tilde{Y},\tilde{Z}|S_t,S_r,S_e,X} &= \mathds{1}_{\{\tilde{Y}=X=\Psi\}} \mathds{1}_{\{\tilde{Z}=0\}},
	\end{align*}
	and $S=S_t=(A,B)$, $Y=(S_r,\tilde{Y})=(T,Q,\Psi)$ and $Z=(S_e,\tilde{Z})=A \oplus B$.	
	
	A valid choice of random variables for \eqref{EQ:Zib}~is
		\footnote{
			To use the original notations of~\cite{Zibaeenejad2015WtSiPublicComm} we identify \(U,V,S_t,S_r,S_e,X,\tilde{Y},\tilde{Z}\) we use, respectively, with \(W,U,S,B,E,X,Y,Z\) from~\cite{Zibaeenejad2015WtSiPublicComm}, where \(C_P = 0\). 
		}
    \begin{enumerate}
		\item \(\Psi \sim \mathsf{Ber}(\frac{1}{2})\) independent of \((A,B,Q)\),
		\item \(U = Z = A \oplus B\),
		\item  \(V = (A,B,\Psi)\),
	\end{enumerate}
	which achieves  $R_\mathsf{Zib}=2$.
	Hence, by showing that the SK capacity of the proposed setup is strictly less than 2, we contradict the achievability of $R_\mathsf{Zib}$ from~\cite[Theorem 1]{Zibaeenejad2015WtSiPublicComm} as a SK rate for this setup. We do so by showing that the \emph{vanishing average error probability} and the \emph{weak secrecy} of the SK, used in the definition of achievability in~\cite{Zibaeenejad2015WtSiPublicComm}, cannot coexist in this setup while a SK rate of 2 is attained.
	
    Consider a sequence of codes \(\{c_n\}_{n\in\mathbb{N}}\) achieving $R_\mathsf{Zib}=2$ for the above setup.
	We have that there exists a sequence \(\{\epsilon_n\}\), with $\lim_{n\to\infty}\epsilon_n=0$, such that
	\begin{subequations}\begin{align}
		H(K_n)&\geq 2n-n\epsilon_n,  \label{EQ:Zib_analize_HL_lower}\\
		H(\Psi^n) &\le n,	\label{EQ:Zib_analize_Hphi}\\
		H(K_n|\Psi^n,S_r^n) &\le n \epsilon_n, \label{EQ:Zib_analize_Fano}\\
		I(K_n;Z^n) &\le n \epsilon_n, \label{EQ:Zib_analize_security}
	\end{align}\label{EQ:Zib_stuff}
	\end{subequations}
	where:
	\begin{enumerate}
		\item[\eqref{EQ:Zib_analize_HL_lower}] follows by the definition of SK rate achievability.
		\item[\eqref{EQ:Zib_analize_Hphi}] is because the alphabet of $\Psi^n$ is of size $2^n$ and since a uniform distribution maximizes discrete entropy.
		\item[\eqref{EQ:Zib_analize_Fano}] is Fano's inequality, following the requirement of vanishing decoding error.
		\item[\eqref{EQ:Zib_analize_security}] is the weak secrecy requirement.
	\end{enumerate}   

    \begin{lemma} \label{LEMMA:Zib_upper_bound}
    For the considered setup, the SK capacity is upper bounded by 2 bits per channel use,
    \begin{equation}
        C^{\mathsf{SK}} \le 2.
    \end{equation}
    \end{lemma}
    Lemma~\ref{LEMMA:Zib_upper_bound} follows because the considered setup, but without an eavesdropper (i.e., when $Z=0$), falls within the framework of the \emph{common randomness} (CR) problem in \emph{Model i} from~\cite{AhlswedeCsiszar1998Part2}.
        
    \begin{IEEEproof}
        Theorem 4.1 in~\cite{AhlswedeCsiszar1998Part2} shows that the CR capacity is upper bounded by
        \begin{equation}
	        C^\mathsf{CR} \le R + I(S;S_r), \label{EQ:Zib_CR_upperBound}
	    \end{equation}
	    where $R$ is the rate of the communication link between the transmitter and the receiver.
	    Evaluating the RHS of \eqref{EQ:Zib_CR_upperBound} with respect to the considered setup shows that it equals 2 (CR bits per channel use).
	    This upper bound remains valid when a security requirement is introduced, since it can only reduce the admissible rates.
    \end{IEEEproof}

    Lemma~\ref{LEMMA:Zib_upper_bound} guarantees the existence of a sequence \(\{\epsilon'_n\}\), with $\lim_{n\to\infty}\epsilon'_n=0$, such that the following condition may be added to the set \eqref{EQ:Zib_stuff}: 
    \begin{equation}
        H(K_n) \le 2n+n\epsilon'_n. \label{EQ:Zib_analize_HL_upper}
    \end{equation}
    
    Another technical lemma we need is stated next. Its proof is omitted due to space limitations. The technique is standard, and the full proof can be found in~\cite[Appendix E]{Bunin_SkSm_Arxiv}.
    \begin{lemma} \label{LEMMA:Zib_main_bound}
        If \eqref{EQ:Zib_analize_HL_lower}-\eqref{EQ:Zib_analize_Fano} hold, then
        \begin{equation}
            H(A^n,B^n|K_n) \le 4n \epsilon_n. \label{EQ:analize_entropyAB_givenL}
        \end{equation}
    \end{lemma}

    Now, combining \eqref{EQ:Zib_analize_HL_upper} and \eqref{EQ:analize_entropyAB_givenL}, we have
	\begin{align*}
		H(K_n|A^n,B^n) &= H(K_n) - H(A^n,B^n) + H(A^n,B^n|K_n) \\
		&\le 2n + n\epsilon'_n - 2n + H(A^n,B^n|K_n) \\
		&\le (4\epsilon_n+\epsilon'_n)n. \numberthis \label{EQ:analize_entropyLgivenAB}
	\end{align*}
	
	Using \eqref{EQ:analize_entropyLgivenAB} we can finally lower bound the conditional information leakage term $I(K_n;\Psi^n,Z^n)$. To do so, first consider
	\begin{align*}
		H(K_n|Z^n) &\le H(K_n,A^n,B^n|Z^n)\\
		&= H(A^n,B^n|Z^n) + H(K_n|A^n,B^n,Z^n)\\
			&\le H(A^n,B^n|Z^n) + H(K_n|A^n,B^n)\\
			&\stackrel{(a)}\le H(A^n,B^n|Z^n) + (4\epsilon_n+\epsilon'_n)n\\
			&\stackrel{(b)}= H(A^n,B^n)-H(Z^n) + (4\epsilon_n+\epsilon'_n)n\\
			&\stackrel{(c)}= (1+4\epsilon_n+\epsilon'_n)n,\numberthis\label{EQ:Zib_analize_HKZ_lower}
			\end{align*}
	where (a) uses \eqref{EQ:analize_entropyLgivenAB}, (b) follows by the chain rule and because $Z^n$ is deterministically defined by $(A^n,B^n)$ and (c) is since $A^n$, $B^n$ and $Z^n=A^n\oplus_n B^n$ are all i.i.d. $\mathsf{Ber}\left(\frac{1}{2}\right)$ sequences, and because $A^n$ and $B^n$ are independent.

    Having \eqref{EQ:Zib_analize_HKZ_lower}, we conclude with
    \begin{align*}
        &I(K_n;Z^n) = H(K_n) - H(K_n|Z^n)\\
        &\ \stackrel{(a)}\ge 2n - n \epsilon_n - (1+4\epsilon_n+\epsilon'_n)n = (1 - 5\epsilon_n - \epsilon'_n)n, \label{EQ:Zib_analize_security_result} \numberthis
    \end{align*}
	where (a) uses \eqref{EQ:Zib_analize_HL_lower} and  \eqref{EQ:Zib_analize_HKZ_lower}. Evidently, \eqref{EQ:Zib_analize_security_result} contradicts \eqref{EQ:Zib_analize_security}.

	\section{Proof of Proposition~\ref{PROP:strict_improv_prbk}} \label{APPEN:proof_example_prbkrn}
        
    Fix $\sigma\in(0,0.5)$ and set
    \begin{subequations}\begin{align}
        &\epsilon=\frac{1}{2} \left[ h\left(\frac{\sigma}{2}\right) - \sigma \right],\\
        &\lambda=h^{-1}(1-\sigma-\epsilon),
    \end{align}\end{subequations}
    where $h:[0,1]\to [0,1]$ and $h^{-1}:[0,1]\to [0,0.5]$ are the binary entropy function and the inverse of its restriction to $[0,0.5]$, respectively. It is readily verified that $\epsilon,\lambda\in(0,0.5)$.	
    By virtue of \eqref{EQ:we_achieve_capacity}, the inner bound from Theorem~\ref{TM:SDWTC_lower_bound} attains the SM capacity, which is given by (see \eqref{EQ:SM_Capacity_Speical_Case})
    \begin{equation}
     C^{\mathsf{SM}} = \min\left\{ C_\mathsf{GP}(W_{Y|S,X}), H(L) \right\},
    \end{equation}
    where $C_\mathsf{GP}(W_{Y|S,X})=\max_{q_{U,X|S}} \big[ I(U;Y) - I(U;S) \big]$ is the GP capacity of the SD channel $W_{Y|S,X}$ with state distribution $W_S$. By the corollary to Theorem 2 from~\cite{HeEl83} we find that $C_\mathsf{GP}(W_{Y|S,X}) = (1-\sigma)(1-\epsilon)$. As $H(L)=1-\sigma-\epsilon<(1-\sigma)(1-\epsilon)$, we obtain
    \footnote{
        The achievability of \eqref{EQ:prabh_exmpl_capacity} may also be verified directly from Theorem~\ref{TM:SDWTC_lower_bound} by substituting $R_K=0$, $U=G$, $V=(U,L)$ and $X \sim \mathsf{Ber}\left(\frac{1}{2}\right)$ independent of $(S,L)$  into \eqref{EQ:SDWTC_alt_lower_bound_prob}.
    }
    \begin{equation}
        C^{\mathsf{SM}} = H(L) = 1 - \sigma - \epsilon = 1 - \frac{1}{2} \left[ \sigma + h \left( \frac{\sigma}{2} \right) \right] \label{EQ:prabh_exmpl_capacity}.
    \end{equation}

    We now show that $R^{\mathsf{SM}}_\mathsf{PER}(\lambda,\epsilon,\sigma)< 1 - \frac{1}{2} \left[ \sigma + h \left( \frac{\sigma}{2} \right) \right]$. Fix a joint distribution to evaluate the region from  \eqref{EQ:compare_prabhakaran} with $R_K=0$, and $S$ and $Y$ replaced with $\tilde{S}=(L,S)$, $\tilde{Y}=(L,Y)$. This distribution factors as
    \begin{align*}
        &q_{L,S,U,V,X,G,E,Y,Z,\tilde{S},\tilde{Y}} \triangleq W_L  W_S q_U q_{V,X|U,S,L}  \mathds{1}_{\{G=g(S,X)\}}\\
        &\times W_E \mathds{1}_{\{Y=y(E,G)\}}\mathds{1}_{\{Z=(S,X)\}} \mathds{1}_{\{\tilde{S}=(L,S)\}\cap\{\tilde{Y}=(L,Y)\}}.\numberthis\label{EQ:prabh_exmpl_jointPMF} 
    \end{align*}
    Note that the independence of $(L,S)$ and $U$ is a restriction on the feasible joint distributions in \eqref{EQ:compare_prabhakaran_LB}.
                
    Now, assume in contradiction that evaluating \eqref{EQ:compare_prabhakaran} with respect to $q$ produces a rate that is at least as high as \eqref{EQ:prabh_exmpl_capacity}. Specifically, assume that             \begin{subequations}\label{EQ:prabh_exmpl_conditions}\
    \begin{equation}
    I(U,V;\tilde{Y}) - I(U,V;\tilde{S}) \geq H(L) \label{EQ:prabh_exmpl_reliabBound}
    \end{equation}
    and
    \begin{equation}
    I(V;\tilde{Y}|U) - I(V;Z|U)\geq H(L). \label{EQ:prabh_exmpl_secrecyBound}
    \end{equation}
    \end{subequations}
        
    Consider the following upper bound on \eqref{EQ:prabh_exmpl_secrecyBound}.
    \begin{align*}        		         
    	&I(V;\tilde{Y}|U) - I(V;Z|U) = I(V;L,Y|U) - I(V;S,X|U)\\
    	&\qquad= I(V;Y|U)+ I(V;L|U,Y) - I(V;S,X|U)\\
        &\qquad= I(V;U,Y) +I(V;L|U,Y) - I(V;U,S,X) \\
        &\qquad\stackrel{(a)}= I(V;L|U,Y) + I(V;U,Y) - I(V;U,S,X,Y)\\
        &\qquad= I(V;L|U,Y) - I(V;S,X|U,Y)\\
        &\qquad= H(L|U,Y) - H(L|U,V,Y) - I(V;S,X|U,Y)\\
        &\qquad\le H(L), \numberthis\label{EQ:pranh_exmpl_sec_bound_2}
    \end{align*}
    where (a) uses the Markov relation $V \mkv (S,U,X) \mkv Y$, which follows because $Y=y\big(E,g(S,X)\big)$ and $E$ is independent of $(S,U,V,X)$ under the distribution from  \eqref{EQ:prabh_exmpl_jointPMF}.
    
    On account of \eqref{EQ:prabh_exmpl_secrecyBound}, the single inequality from  \eqref{EQ:pranh_exmpl_sec_bound_2} must hold with equality. For this to happen, all the following arguments must hold.
    \begin{enumerate}
        \item The conditioning is removed from the first (positive) term, i.e., 
            $H(L)=H(L|U,Y)$.
        This implies that $L$ is independent of $(U,Y)$.
        \ \\
        \item The second (negative) term is zero, i.e.,
        \begin{align*}
            0&=H(L|U,V,Y) \stackrel{(a)}=H(L|U,V,Y,E)\\
            &=(1-\epsilon)\cdot H(L|U,V,Y,E=0)\\
            &\qquad+\epsilon\cdot H(L|U,V,Y,E=1),\numberthis
        \end{align*}
    	where (a) is because $E$ is deterministically defined by $Y$. Now, since $\epsilon>0$, we have that $H(L|U,V,Y,E=1)=0$. Observing that conditioned on $\{E=1\}$, $Y=?$ is a constant, we further deduce
        \begin{align*}
            H(L|U,V,Y,E=1) &= H(L|U,V,E=1)\\
            &\stackrel{(a)}= H(L|U,V) = 0, \numberthis\label{EQ:last_equality}
        \end{align*}
        where (a) relies on the independence of $E$ and $(L,U,V)$. The last equality in \eqref{EQ:last_equality} implies that there exists a (deterministic) function $\ell:\mathcal{U}\times\mathcal{V}\to\mathcal{L}$ such that $L=\ell(U,V)$.
        \ \\		
        \item \label{ENUM:example_prbhkrn_markov} Expanding the third (negative) term with respect to $E$ in a similar manner to that presented in the above 2nd point, we obtain
    	\begin{align*}
            I(V;S,X|U,Y,E=1) &= I(V;S,X|U,E=1) \\
            &= I(V;S,X|U) = 0, \numberthis
        \end{align*}
        which establishes $V \mkv U \mkv (S,X)$ as a Markov chain. 
    \end{enumerate}

    Since $S$ and $U$ are independent under $q$ from \eqref{EQ:prabh_exmpl_jointPMF}, the Markov relation from point~\ref{ENUM:example_prbhkrn_markov}) further implies that $S$ is independent of the pair $(U,V)$. Observe that this effectively means that the inability of the scheme from~\cite[Theorem 1]{Prabhakaran_SKSM2012} to support GP coding in the inner layer implies that GP coding is not supported at all.
        	
    We proceed to analyze \eqref{EQ:prabh_exmpl_reliabBound} under the above deductions. Consider
	\begin{align*}
		&I(U,V;\tilde{Y}) - I(U,V;\tilde{S}) = I(U,V;L,Y) - I(U,V;L,S)\\
		&\qquad= I(U,V;Y|L) - I(U,V;S|L)\\
		&\qquad\le I(U,V,L;Y)
		\stackrel{(a)}\le I(U,V,L;G)
		\stackrel{(b)}= I(U,V;G), \numberthis\label{EQ:pranh_exmpl_rel_bound_2}
	\end{align*}
    where (a) follows by the Data Processing Inequality (see, e.g.,~\cite[Section 2.8]{CovThom06}) and since $(L,U,V) \mkv G \mkv Y$ forms a Markov chain, while (b) is because $L=\ell(U,V)$. 

    Define $T=(U,V)$ and observe that $T$ is independent of $S$ (since the pair $(U,V)$ is) and that $T \mkv (S,X) \mkv G$ forms a Markov chain (since $G=g(S,X)$). We further upper bound the RHS of \eqref{EQ:pranh_exmpl_rel_bound_2} with $T=(U,V)$ by maximizing it over all conditional distributions that satisfy $q_{T,X|S}=q_T  q_{X|S,T}$. We thus have
    \begin{equation}
        I(U,V;\tilde{Y}) - I(U,V;\tilde{S}) \le I(T;G)
        \le \max_{q_T q_{X|S,T}} I(T;G). \label{EQ:pranh_exmpl_rel_bound_3}
    \end{equation}

    The expression on the RHS of  \eqref{EQ:pranh_exmpl_rel_bound_3} is the capacity of the MSAF with causal encoder knowledge of the state sequence (cf., e.g.,~\cite[p.5469]{Jafar2006}). However, the causal CSI is useless for the MSAF encoder, as demonstrated in Section V-A of~\cite{Jafar2006}. Omitting the availability of any CSI from the MSAF encoder, the channel is equivalent to a binary symmetric channel with flip probability $\frac{\sigma}{2}$ (see \eqref{EQ:MSAF_def}), whose capacity equals $1-h \left( \frac{\sigma}{2} \right)$. 

    We conclude with
    \begin{align*}
        &I(U,V;\tilde{Y}) - I(U,V;\tilde{S}) \le \max_{q_T q_{X|TS}} I(T;G)\\
        &\qquad= 1 - h \left( \frac{\sigma}{2} \right) 
        \stackrel{(a)}< 1 - \frac{1}{2} \left[ \sigma + h \left( \frac{\sigma}{2} \right) \right]
        =H(L),\numberthis\label{EQ:pranh_exmpl_rel_bound_4}
    \end{align*}
    where (a) is because  $\sigma<h(\frac{\sigma}{2})$ for any $\sigma\in(0,0.5)$. This is a contradiction to \eqref{EQ:prabh_exmpl_reliabBound}.

\bibliographystyle{IEEEtran}
\bibliography{ref}

\begin{thebibliography}{10}
\providecommand{\url}[1]{#1}
\csname url@samestyle\endcsname
\providecommand{\newblock}{\relax}
\providecommand{\bibinfo}[2]{#2}
\providecommand{\BIBentrySTDinterwordspacing}{\spaceskip=0pt\relax}
\providecommand{\BIBentryALTinterwordstretchfactor}{4}
\providecommand{\BIBentryALTinterwordspacing}{\spaceskip=\fontdimen2\font plus
\BIBentryALTinterwordstretchfactor\fontdimen3\font minus
  \fontdimen4\font\relax}
\providecommand{\BIBforeignlanguage}[2]{{%
\expandafter\ifx\csname l@#1\endcsname\relax
\typeout{** WARNING: IEEEtran.bst: No hyphenation pattern has been}%
\typeout{** loaded for the language `#1'. Using the pattern for}%
\typeout{** the default language instead.}%
\else
\language=\csname l@#1\endcsname
\fi
#2}}
\providecommand{\BIBdecl}{\relax}
\BIBdecl

\bibitem{Bunin_WCS2017}
A.~Bunin, Z.~Goldfeld, H.~H. Permuter, S.~Shamai~(Shitz), P.~Cuff, and
  P.~Piantanida, ``Semantically-secured message-key trade-off over wiretap
  channels with random parameters,'' in \emph{Proceedings of the 2nd Workshop
  on Communication Security: Cryptography and Physical Layer Security}.\hskip
  1em plus 0.5em minus 0.4em\relax Springer International Publishing, 2018, pp.
  33--48.

\bibitem{Bunin_SkSm_Arxiv}
------, ``Semantically-secured message-key trade-off over wiretap channels with
  random parameters,'' \emph{ArXiv preprint}, Aug 2017, available at
  https://arxiv.org/abs/1708.04283v1.

\bibitem{Bloch_Barros_Secrecy_Book2011}
M.~Bloch and J.~Barros, \emph{Physical-Layer Security: From Information Theory
  to Security Engineering}.\hskip 1em plus 0.5em minus 0.4em\relax Cambridge,
  UK: Cambridge Univ. Press, Oct. 2011.

\bibitem{Liu_PhySecurity_Tutorial2017}
Y.~Liu, H.~H. Chen, and L.~Wang, ``Physical layer security for next generation
  wireless networks: Theories, technologies, and challenges,'' \emph{IEEE
  Commun. Surv. Tut.}, vol.~19, no.~1, pp. 347--376, First quarter 2017.

\bibitem{Zeng2015_PLS_SK_Survey}
K.~Zeng, ``Physical layer key generation in wireless networks: challenges and
  opportunities,'' \emph{IEEE Commun. Mag.}, vol.~53, no.~6, pp. 33--39, June
  2015.

\bibitem{Wyner_Wiretap1975}
A.~D. Wyner, ``The wire-tap channel,'' \emph{Bell Sys. Techn.}, vol.~54, no.~8,
  pp. 1355--1387, Oct. 1975.

\bibitem{Csiszar_Korner_BCconfidential1978}
I.~Csisz{\'a}r and J.~K{\"o}rner, ``Broadcast channels with confidential
  messages,'' \emph{IEEE Trans. Inf. Theory}, vol.~24, no.~3, pp. 339--348, May
  1978.

\bibitem{Wyner_Common_Information1975}
A.~D. Wyner, ``The common information of two dependent random variables,''
  \emph{IEEE Trans. Inf. Theory}, vol.~21, no.~2, pp. 163--179, Mar. 1975.

\bibitem{Han_Verdu_Resolvability1993}
T.~Han and S.Verd{\'u}, ``Approximation theory of output statistics,''
  \emph{IEEE Trans. Inf. Theory}, vol.~39, no.~3, pp. 752--772, May 1993.

\bibitem{Kramer_resolvability2013}
J.~Hou and G.~Kramer, ``Informational divergence approximations to product
  distributions,'' in \emph{Proc. 13th Canadian Workshop Inf. Theory (CWIT)},
  Toronto, Ontario, Canada, Jun. 2013.

\bibitem{Goldfeld_WTCII2016}
Z.~Goldfeld, P.~Cuff, and H.~H. Permuter, ``Semantic-security capacity for
  wiretap channels of type {II},'' \emph{IEEE Trans. Inf. Theory}, vol.~62,
  no.~7, pp. 3863--3879, Jul. 2016.

\bibitem{Goldfeld_AVWTC2016}
------, ``Arbitrarily varying wiretap channels with type constrained states,''
  \emph{IEEE Trans. Inf. Theory}, vol.~62, no.~12, pp. 7216--7244, Dec. 2016.

\bibitem{BastaniParizi2017SecrecyExponent}
M.~B. Parizi, E.~Telatar, and N.~Merhav, ``Exact random coding secrecy
  exponents for the wiretap channel,'' \emph{IEEE Trans. Inf. Theory}, vol.~63,
  no.~1, pp. 509--531, Jan 2017.

\bibitem{YagliCuff2018ISIT}
S.~Yagli and P.~Cuff, ``Exact soft-covering exponent,'' in \emph{2018 IEEE Int.
  Symp. Inf. Theory (ISIT)}, June 2018, pp. 1680--1684.

\bibitem{Maurer1993}
U.~M. Maurer, ``Secret key agreement by public discussion from common
  information,'' \emph{IEEE Trans. Inf. Theory}, vol.~39, no.~3, pp. 733--742,
  May 1993.

\bibitem{AhlswedeCsiszar1993Part1}
R.~Ahlswede and I.~Csisz{\'a}r, ``Common randomness in information theory and
  cryptography. part i: secret sharing,'' \emph{IEEE Trans. Inf. Theory},
  vol.~39, no.~4, pp. 1121--1132, Jul. 1993.

\bibitem{csiszarNarayan2000CrHelper}
I.~Csisz{\'a}r and P.~Narayan, ``Common randomness and secret key generation
  with a helper,'' \emph{IEEE Trans. Inf. Theory}, vol.~46, no.~2, pp.
  344--366, Aug. 2000.

\bibitem{WynZiv76SideInformationDecoder}
A.~D. Wyner and J.~Ziv, ``The rate-distortion function for source coding with
  side information at the decoder,'' \emph{IEEE Trans. Inf. Theory}, vol.~1,
  pp. 1--10, 1976.

\bibitem{Gelfand_Pinsker}
S.~I. Gelfand and M.~S. Pinsker, ``Coding for channel with random parameters,''
  \emph{Problemy Pered. Inform. (Problems of Inf. Trans.)}, vol.~9, no.~1, pp.
  19--31, 1980.

\bibitem{SDWTC_Chen_HanVinck2006}
Y.~Chen and A.~J.~H. Vinck, ``Wiretap channel with side information,''
  \emph{IEEE Trans. Inf. Theory}, vol.~54, no.~1, pp. 395--402, Jan. 2008.

\bibitem{Goldfeld_SDWTC2016}
Z.~Goldfeld, P.~Cuff, and H.~H. Permuter, ``Wiretap channel with random states
  non-causally available at the encoder,'' \emph{ArXiv preprint}, 2016,
  available at https://arxiv.org/abs/1608.00743v2.

\bibitem{Fujita2016SDWTC_causalCSI}
H.~Fujita, ``On the secrecy capacity of wiretap channels with side information
  at the transmitter,'' \emph{IEEE Trans. Inf. Forensic Secur.}, vol.~11,
  no.~11, pp. 2441--2452, Nov 2016.

\bibitem{HanSaaki17SDWTC_causalCSI}
T.~S. Han and M.~Sasaki, ``Wiretap channels with causal state information:
  Strong secrecy,'' \emph{ArXiv preprint}, Aug 2017, available at
  https://arxiv.org/abs/1708.00422.

\bibitem{Khisti_Key_Agreement2011}
A.~Khisti, S.~N. Diggavi, and G.~W. Wornell, ``Secret-key agreement with
  channel state information at the transmitter,'' \emph{IEEE Trans. Inf.
  Forensic Secur.}, vol.~6, no.~3, pp. 672--681, Mar. 2011.

\bibitem{Bassi2016secretKey}
G.~Bassi, P.~Piantanida, and S.~Shamai~(Shitz), ``Secret key generation over
  noisy channels with common randomness,'' \emph{ArXiv preprint}, Sep. 2016,
  available at https://arxiv.org/abs/1609.08330.

\bibitem{Prabhakaran_SKSM2012}
V.~Prabhakaran, K.~Eswaran, and K.~Ramchandran, ``Secrecy via sources and
  channels,'' \emph{IEEE Trans. Inf. Theory}, vol.~85, no.~11, pp. 6747--6765,
  Nov. 2012.

\bibitem{Cuff_Song_Likelihood2016}
E.~Song, P.~Cuff, and V.~Poor, ``The likelihood encoder for lossy
  compression,'' \emph{IEEE Trans. Inf. Theory}, vol.~62, no.~4, pp.
  1836--1849, Apr. 2016.

\bibitem{Vardy_Semantic_WTC2012}
M.~Bellare, S.~Tessaro, and A.~Vardy, ``A cryptographic treatment of the
  wiretap channel,'' in \emph{Proc. Adv. Crypto. (CRYPTO 2012)}, Santa Barbara,
  CA, USA, Aug. 2012.

\bibitem{Zibaeenejad2015WtSiPublicComm}
A.~Zibaeenejad, ``Key generation over wiretap models with non-causal side
  information,'' \emph{IEEE Trans. Inf. Forensic Secur.}, vol.~10, no.~7, pp.
  1456--1471, July 2015.

\bibitem{Eggleston_Convexity1958}
H.~G. Eggleston, \emph{Convexity}.\hskip 1em plus 0.5em minus 0.4em\relax
  Cambridge University Press, 1958.

\bibitem{Shamai_ITSecurity2009}
Y.~Liang, H.~V. Poor, and S.~Shamai, ``Information theoretic security,''
  \emph{Foundations and Trends{\textregistered} in Commun. and Inf. Theory},
  vol.~5, no. 4-5, pp. 355--580, 2009.

\bibitem{SDWTC_2Sided_Liu2007}
W.~Liu and B.~Chen, ``Wiretap channel with two-sided state information,'' in
  \emph{Proc. 41st Asilomar Conf. Signals, Syst. Comp}, Pacific Grove, CA, US,
  Nov. 2007, p. 893–897.

\bibitem{SDWTC_Chia_ElGamal2012}
Y.-K. Chia and A.~E. Gamal, ``Wiretap channel with causal state information,''
  \emph{IEEE Trans. Inf. Theory}, vol.~58, no.~5, pp. 2838--2849, May 2012.

\bibitem{Khisti_Key_Generation2012}
A.~Khisti, S.~N. Diggavi, and G.~W. Wornell, ``Secret-key generation using
  correlated sources and channels,'' \emph{IEEE Trans. Inf. Theory}, vol.~58,
  no.~2, pp. 652--670, Feb. 2012.

\bibitem{Shannon58}
C.~E. Shannon, ``Channels with side information at the transmitter,'' \emph{IBM
  J. Res. Devel.}, vol.~2, no.~4, pp. 289--293, Oct. 1958.

\bibitem{Tsybakov_Memory_stuckat1974}
A.~V. Kuznetsov and B.~S. Tsybakov, ``Coding in a memory with defective
  cells,'' \emph{Problemy Pered. Inform. (Problems of Inf. Trans.)}, vol.~10,
  no.~2, pp. 52--60, 1974.

\bibitem{CovThom06}
T.~M. Cover and J.~A. Thomas, \emph{Elements of Information Theory},
  2nd~ed.\hskip 1em plus 0.5em minus 0.4em\relax New-York: Wiley, 2006.

\bibitem{Han_Vinck_SDWTC_Actions2013}
B.~Dai, A.~J.~H. Vinck, Y.~Luo, and X.~Tang, ``Wiretap channel with
  action-dependent channel state information,'' \emph{Entropy}, vol.~15, pp.
  445--473, 2013.

\bibitem{AhlswedeCsiszar1998Part2}
R.~Ahlswede and I.~Csisz{\'a}r, ``Common randomness in information theory and
  cryptography. ii. cr capacity,'' \emph{Information Theory, IEEE Transactions
  on}, vol.~44, no.~1, pp. 225--240, 1998.

\bibitem{HeEl83}
C.~Heegaard and A.~E. Gamal, ``On the capasity of computer memories with
  defects,'' \emph{IEEE Trans. Inf. Theory}, vol. IT-29, pp. 731--739, Sept.
  1983.

\bibitem{Jafar2006}
S.~A. Jafar, ``Channel capacity with causal and noncaudal side information - a
  unified view,'' \emph{IEEE Trans. Inform. Theory}, vol.~52, no.~12, pp.
  5468--5474, Dec. 2006.

\end{thebibliography}

\end{document}